\documentclass{aa}
\usepackage{amsmath,bm}
\usepackage{graphicx}
\usepackage{enumerate}
\usepackage{bm}
\usepackage{times}
\usepackage{natbib}
\bibpunct{(}{)}{;}{a}{}{,}

\graphicspath{{./fig/}{./png/}}

\newcommand{\EQ}{\begin{equation}}
\newcommand{\EE}{\end{equation}}
\newcommand{\EQA}{\begin{eqnarray}}
\newcommand{\EEA}{\end{eqnarray}}
\newcommand{\brac}[1]{\langle #1 \rangle}
\newcommand{\pd}{\partial}

\newcommand{\DIV}{\vec{\nabla} \cdot }

\newcommand{\mean}[1]{\overline{#1}}
\newcommand{\meanv}[1]{\overline{\bm #1}}

\newcommand{\etat}{\eta_{\rm t}}

\newcommand{\urms}{u_{\rm rms}}

\newcommand{\brms}{B_{\rm rms}}
\newcommand{\Beq}{B_{\rm eq}}
\newcommand{\Ma}{\rm Ma}

\newcommand{\kef}{k_{\rm f}}

\newcommand{\Sh}{{\rm Sh}}
\newcommand{\Pm}{{\rm Pm}}
\newcommand{\Rm}{{{\rm Rm}}}
\newcommand{\Pra}{{\rm Pr}}
\newcommand{\Ra}{{\rm Ra}}

\newcommand{\Rey}{{\rm Re}}

{}
\newcommand{\memf}{\overline{\mbox{\boldmath ${\cal E}$}}{}}{}
\def\onethird{{\textstyle{1\over3}}}
\def\onehalf{{\textstyle{1\over2}}}

\begin{document}

\authorrunning{Guerrero \& K\"apyl\"a}
\titlerunning{Dynamo action due to localized shear below a convection zone}

   \title{Dynamo action and magnetic buoyancy in convection
     simulations with vertical shear}

   \author{
          G. Guerrero
          \inst{1}
          \and
          P. J. K\"apyl\"a
	  \inst{1,2}
        }

   \offprints{G. Guerrero\\
          \email{guerrero@nordita.org}
	  }

          \institute{NORDITA, Roslagstullsbacken
            23, SE-10691 Stockholm, Sweden
            \and Department of Physics, Gustaf H\"allstr\"omin katu 2a 
            (PO Box 64), FI-00014 University of Helsinki, Finland}

   \date{ Received / Accepted }

   \abstract
   {A hypothesis for sunspot formation is the buoyant
     emergence of magnetic flux tubes created by the strong radial
     shear at the tachocline.
     In this scenario, the magnetic 
     field has to exceed a threshold value before it becomes buoyant
     and emerges through the whole convection zone.}
   {We follow the evolution of a random seed magnetic field with
     the aim of study under what conditions it is possible to
     excite the dynamo instability and whether 
     the dynamo generated magnetic field becomes buoyantly unstable and
     emerges to the surface as expected in the flux-tube context.}
   {We perform numerical simulations of compressible turbulent
     convection that include a vertical
     shear layer. Like the solar tachocline, the shear is located at 
     the interface  between convective and stable layers.}
   {We find that shear and convection are able to amplify 
     the initial magnetic field and form large-scale 
     elongated magnetic structures. The magnetic field strength
     depends on several parameters such as the shear amplitude,
     the thickness and location of the shear layer, and the
     magnetic Reynolds number ($\Rm$).
     Models with deeper and thicker
     tachoclines allow longer storage and are more favorable 
     for generating a mean magnetic
     field. Models with higher $\Rm$ grow faster
     but saturate at slightly lower levels.
     Whenever the toroidal magnetic field reaches amplitudes 
     greater a threshold value which is close to the equipartition 
     value, it becomes buoyant and
     rises into the convection zone where it expands and forms mushroom
     shape structures. Some events of emergence, i.e. those with the 
     largest
     amplitudes of the initial field, are able to reach the very uppermost
     layers of the domain. These episodes are able to modify the
     convective pattern forming either broader convection cells or
     convective eddies elongated in the direction of the
     field. However, in none of these events the field preserves its
     initial structure. The back-reaction 
     of the magnetic field on the fluid is also observed in lower
     values of the turbulent velocity and in perturbations of approximately 
     three per cent on the shear profile.}
   {The results indicate that buoyancy is a common phenomena when the
     magnetic field is amplified 
through dynamo action
in a narrow layer. It is, however,
     very hard for the field to rise up to the surface without losing
     its initial coherence.}


   \maketitle


\section{Introduction}
\label{sec:int}

Sunspots appear at the solar surface following a 11-year cycle.
They reveal the presence of strong magnetic fields in the solar
interior, and suggest the existence of a dynamo process governing its
evolution. However,  
the process by which sunspots are formed is still unknown. 
At the solar surface, sunspots are observed as bipolar patches of
radial magnetic field. This intuitively suggest that they are formed
by the emergence of  horizontal concentrations of magnetic field
lines, often called magnetic flux tubes. Since Parker's original
proposal of magnetic buoyancy \citep{P55}, the model has evolved
through the thin flux tube approximation \citep{Spruit_81}, to
numerical simulations of the emergence of 3D magnetic flux tubes
\citep{Fan_08}. The buoyant rise of flux tubes from the tachocline up
to the surface is currently the most widely accepted mechanism of
sunspot formation. During the last three decades
much work has been done in order to understand the buoyancy
phenomena and to reconcile the results of flux tube models with 
phenomenological
sunspots rules such as the Joy's law or the topological
difference between the two spots in a pair. In spite of the fact 
that these basic observations have been qualitatively reproduced 
by the flux tube models, a set of complications have questioned the
feasibility of this scenario. These may be summarized as follows:

\begin{enumerate}
\item
Magnetic flux tubes at the base of the convection zone should have a
strength of 
around $10^5$ G in order to become unstable and then cross
the entire convection zone up to the surface \citep{caligari+etal_95}.
This has been considered a problem since the magnetic energy density corresponding to such field is one to two orders of magnitude larger than the kinetic energy of the turbulent motions (the equipartition energy). 
Note, however, that in the presence of shear this is not any longer an upper limit for the amplitude of the magnetic field \cite[see e.g.][]{kapyla+bran_09}. 
\item
Sunspots are coherent magnetic structures which means that the flux
tubes should preserve their integrity while they rise through the entire
convection zone. However, this region is highly turbulent and 
stratified, spanning more than 20 pressure scale heights, so in 
addition to the $10^5$ G
strength, the tubes must have a certain amount of twist in order to
resist the turbulent diffusion 
\cite[see e.g.][]{Emonet+Moreno_98,fan+etal_03}. The current 
results do not conclusively yield the amount of twist required for the
tubes to rise coherently up to the surface. 
On one hand, 2D simulations require a larger twist in
order to prevent the fragmentation of the flux tube in two tubes
due to the vorticity generated by the buoyant rise
\citep{schussler_79,moreno+emonet_96,Longcope+etal_96,Emonet+Moreno_98,Fan+etal_98}.
On the other hand, rising 3D flux tubes require less twist thanks to
the tension forces due to the longitudinal curvature of the
tube \citep{Fan_01}. Three dimensional simulations in spherical
coordinates require fine tuning of the initial
twist for the tubes to emerge to the surface with the observed
tilt \citep{Fan_08}. 
\item
MHD simulations of convection in Cartesian coordinates have been 
able to produce large-scale magnetic fields through $\alpha^2$
\citep{kapyla+etal_09} and $\alpha\Omega$ \citep{kkb08}
dynamo action. However, the magnetic fields generated  on these
simulations are more homogeneously distributed in space rather than in
the form of isolated magnetic structures. To the date, however, no
direct simulations have been able to spontaneously form sunspot-like
magnetic structures.
\item
Furthermore, instead of showing signs of buoyant rise and emergence of the
magnetic field, the 3D dynamo simulations above have shown that the
magnetic field is pumped down by convective downflows and tends to
remain in the stable layer
\citep[see also][]{Tobias+etal_98,Tobias+etal_01,OSBR02}. 
\cite{fan+etal_03} studied the 
evolution of an isolated flux tube in a turbulent convection zone. They
found that coherent rise is possible as far as the magnetic buoyant 
force overcomes the
hydrodynamic force from convection,  i.e., obeying the condition
$B_0 > (H_{\rm P}/a)^{1/2} \Beq$, where $B_0$ is the initial magnetic field
strength, $H_{\rm P}$ is the pressure scale height, and $a$ is the radius of
the tube. In their case where $(H_{\rm P}/a)^{1/2} \approx 3$,
magnetic flux tubes 
with $B_0 \ge 3 \Beq$ are able to rise coherently without being
affected by convection. Similar results were found in a similar setup
in spherical geometry \citep{jouve+brun_09}. However, this does not
address the issue with the pumping since, firstly, a strong magnetic
flux tube is imposed on the convective layer, and secondly, magnetic
pumping is an 
effect related with the gradient of turbulence intensity
\citep{Kit+Rud_92}, which is not present in these simulations. 
 \end{enumerate}

In addition to the thin flux tube approximation and simulations of rising
flux tubes on stratified atmospheres, other recent attempts have
been made in order to simulate the formation of a magnetic layer
through the interaction of an imposed shear in convectively stable
\citep{vasil+etal_08} and unstable \citep{silvers+etal_09a}
atmospheres, with an imposed vertical magnetic field. 
They have found that,
unlike in the cases of imposed toroidal magnetic layers, buoyancy 
instability is
harder to excite when the magnetic field is generated by the shear. 
More recently \cite{silvers+etal_09b}, have
found, with a similar setup, that the buoyancy may be
favored by the presence of double-diffusive instabilities (these
in turn depend on the ratio 
between the thermal and magnetic diffusivities, $\chi/\eta$, often
known as the inverse Roberts number). Later independent study of
\cite{chatterjee_11} has confirmed this result.
However, in most of the current models, the presence of
stratified turbulence, self-consistent generation of the magnetic
field, or both, are omitted.

In view of the above mentioned issues, other mechanisms have been
proposed in order to explain sunspots. 
These are related to instabilities due to the presence of a
diffuse large-scale magnetic field in a highly stratified turbulent
medium \citep{Kleo+Roga_94,roga+kleo_07,bran+etal_10b,bran+etal_10a}.
Mean-field models using this mechanism are able to produce strong 
flux concentrations but this has not yet been achieved in direct
numerical simulations.

Here we present numerical simulations of compressible turbulent
convection 
with an imposed radial shear flow located in a sub-adiabatic layer
beneath the convective region. For numerical reasons,
explained below, we do not include rotation in our setup, i.e.\ the
turbulence is not helical, and an $\alpha\Omega$ dynamo is not
expected. Nevertheless, recent studies have shown that mean-field
dynamo action is possible due to non-helical turbulence and
shear \citep{bran05,yousef_etal_08b,yousef_etal_08a,BRRK08}. The
nature of  this dynamo is not yet entirely clear and may be attributed
to the so called shear-current effect \citep{rog+kleo_03,rog+kleo_04}
or to the  incoherent, stochastic, $\alpha$-effect
\citep{vish+bran_97}. 
According to \citep{BRRK08} the latter explanation is consistent
with the turbulent transport coefficients.

Based on these results, we expect the development of a mean field
magnetic field, i.e. dynamo action, with a system that mimics, as far
as possible,  the conditions of the solar interior, specifically in
the lower part of the convection zone and the tachocline.  As the
shear is localized in a very narrow layer, we also expect the
formation of a magnetic layer and the subsequent buoyancy of the
magnetic fields.  

A similar setup was studied recently by \cite{Tobias+etal_08}. They
reported the appearance of elongated stripes of magnetic field in the
direction of the shear. However, since they considered the
Boussinesq approximation in their simulations, no buoyancy was
observed. They also do not report the presence of a large scale
dynamo. 

Two important features 
distinguish the simulations presented here from previous studies in
the context of flux tube formation and emergence. Firstly, we
consider a highly stratified domain with $\approx8$ scale heights in pressure  and $\approx6$ scale heights in density.
Secondly,  we do not impose a background radial magnetic
field but allow the self-consistent development of the field from a
initial random seed.  

Even though this is a complicated setup where it is difficult to 
analyze the different processes occurring independently, we believe
that these simulations may give us some light on the current
paradigm of sunspot formation. There are several important issues
that we want to address with the following simulations. (1) What are
requirements for dynamo action in the present setup? This
includes the dependence of the dynamo excitation on several parameters
such as the amplitude of the shear, thickness and location of the
shear layer, and the aspect ratio of the box. (2) What is the
resulting configuration of the magnetic 
field? In particular, whether the field is predominantly in small or 
large scales, and whether
it is organized in the form of a magnetic layer or isolated
magnetic flux tubes. (3) Is the buoyancy instability \citep{P55}
operating on these magnetic structures? If yes, (4) how it
depends on the parameters listed above? (5) Is it possible to have 
magnetic structures strong enough to emerge from the shear layer
to the surface without being affected by the turbulent convective
motions? (6) Finally, it is important to study how these strong
structures back-react on the fluid motions, including the shear
profile as well as the convective pattern. 

Another important issue that may be addressed in this context is the
mechanism that triggers the dynamo instability. With the recent
developments on  the test-field method, it is possibly to compute the
dynamo transport coefficients and have a better understanding on the
underlying mechanism.  

We have organized this paper as follow: in Sect.~\ref{sec:model} we
provide the details of the numerical model, in Sect.~\ref{sec:results}
we describe our results. We summarize and conclude in
Sect.~\ref{sec:conclusions}.

\section{The model}
\label{sec:model}
Our model setup is similar to that used by e.g.\ 
\cite{bjnrst96} and \cite{kkb08}. A rectangular
portion of a star is modeled by a box whose dimensions are $(L_x,
L_y, L_z) = (4,4,2)d$, where $d$ is the depth of the convectively
unstable layer, which is also used as the unit of length. The box is
divided into three layers, an upper cooling layer, a convectively
unstable layer, and a stable overshoot layer (see below). The
following set of equations for compressible magnetohydrodynamics is
being solved:
\begin{equation}
\frac{\pd \bm A}{\pd t} = \bm{U}\times\bm{B}-\mu_0\eta {\bm J}, \label{equ:AA}
\end{equation}
\begin{equation}
\frac{D \ln \rho}{Dt} = -\DIV{\bm U},
\end{equation}
\begin{equation}
 \frac{D \bm U}{Dt} = -\frac{1}{\rho}{\bm \nabla}p + {\bm g} +
 \frac{1}{\rho} \bm{J} \times {\bm B} + \frac{1}{\rho} \bm{\nabla}
 \cdot 2 \nu \rho \mbox{\boldmath ${\sf S}$}
 -\frac{{\bm U}-\meanv{U}^{(0)}}{\tau_{\rm f}}, \label{equ:UU} 
\end{equation}
\begin{equation}
T \frac{D s}{Dt} = \frac{1}{\rho}
 \bm{\nabla} \cdot K \bm{\nabla}T + 2 \nu \mbox{\boldmath ${\sf S}$}^2
 + \frac{\mu_0\eta}{\rho} \bm{J}^2 - \Gamma, \label{equ:ene} 
\end{equation}
where $D/Dt = \pd/\pd t + \bm{U} \cdot
\bm{\nabla}$ is the total time derivative. 
$\bm{A}$ is the magnetic vector potential, $\bm{B} =
\bm{\nabla} \times \bm{A}$ is the magnetic field, and $\bm{J}
=\bm{\nabla} \times \bm{B}/\mu_0$ is the current density, $\mu_0$ is
the magnetic permeability, $\eta$ and $\nu$ are the magnetic
diffusivity and kinematic viscosity, respectively, $T$ is the 
temperature, $s$ is the specific entropy, $K$ is the heat
conductivity, $\rho$ is the density, $\bm{U}$ is the velocity, and $\bm{g} =
-g\hat{\bm{z}}$ is the gravitational acceleration. The fluid obeys an
ideal gas law $p=\rho e (\gamma-1)$, where $p$ and $e$ are the
pressure and internal energy, respectively, and $\gamma = c_{\rm
  P}/c_{\rm V} = 5/3$ is the ratio of specific heats at constant
pressure and volume, respectively. The specific internal energy per
unit mass is related to the temperature via $e=c_{\rm V} T$. The rate
of strain tensor $\mbox{\boldmath ${\sf S}$}$ is given by
\begin{equation}
{\sf S}_{ij} = \onehalf (U_{i,j}+U_{j,i}) - \onethird \delta_{ij} \DIV \bm{U},
\end{equation}
where the commas denote derivatives.
The last term of Eq.~(\ref{equ:UU}) relaxes the horizontally averaged
mean velocity $\meanv{U}$ towards a target profile $\meanv{U}^{(0)}$
where $\tau_{\rm f}=2\sqrt(dg)$ is a relaxation time scale. 
In a non-shearing case this value corresponds to $\approx1/2\tau_{\rm turn}$,  where $\tau_{\rm turn}=(\urms \kef)^{-1}$  gives an estimate of the convective turnover time of the turnover time.
The target profile is chosen so that it mimics the radial shear in the solar
tachocline. More details of the shear profiles are given in
Sect.~(\ref{sec:shear}). 

The last term of Eq.~(\ref{equ:ene}) describes cooling at the top of
the domain according to
\begin{equation}
\Gamma_{\rm cool} = \Gamma_0 f(z) \left(\frac{T-T_4}{T_4}\right),
\end{equation}
where $f(z)$ is a profile function equal to unity in $z>z_3$ and
smoothly connecting to zero below, and $\Gamma_0$ is a cooling
luminosity chosen so that the sound speed in the uppermost layer
relaxes toward $T_4=T(z=z_4)$.

The positions of the bottom of the box, bottom and top of the
convectively unstable layer, and the top of the box, respectively,
are given by $(z_1, z_2, z_3, z_4) = (-0.8, 0, 1, 1.2)d$. Initially
the stratification is piecewise polytropic with polytropic indices
$(m_1, m_2, m_3) = (6, 1, 1)$, which leads to a convectively unstable
layer above a stable layer at the bottom of the domain and an
isothermal cooling layer at the top.
\begin{figure}[htb]
\centering
\includegraphics[width=0.99\columnwidth]{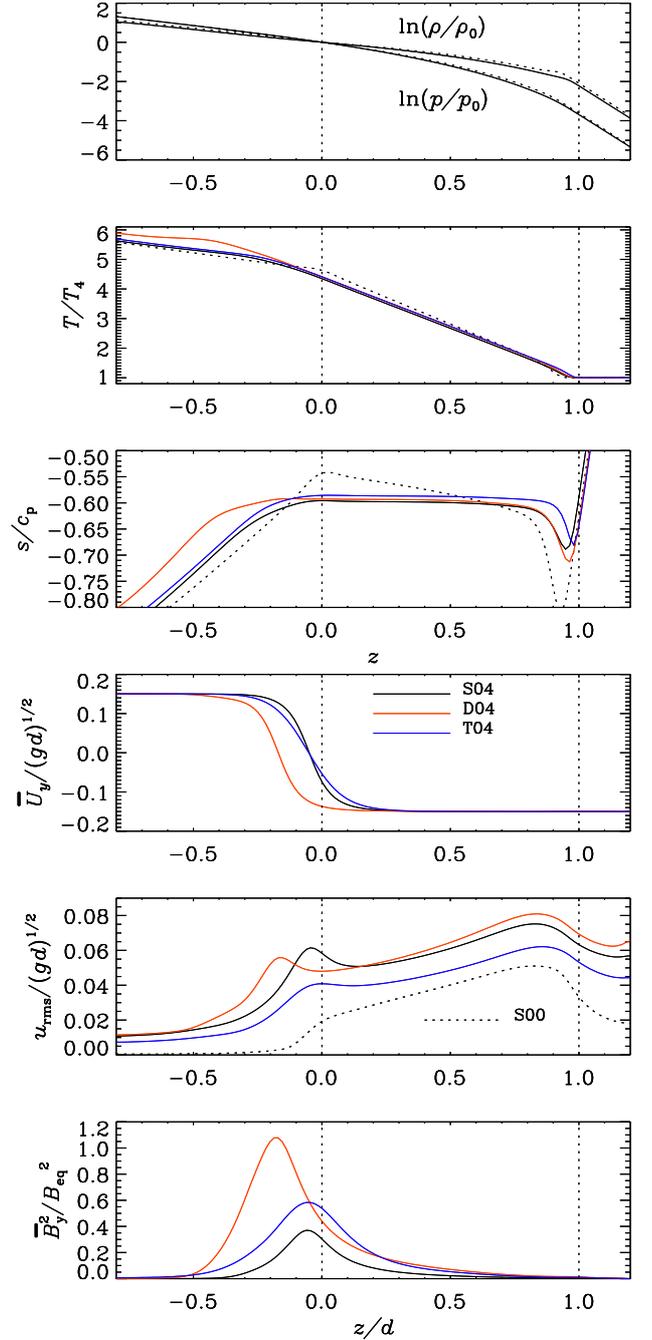} 
\caption{Vertical profiles of mean  density,
  pressure, temperature, specific entropy and azimuthal velocity ($\mean{U}_y$)
  in the initial state (solid lines) and the in the thermally saturated 
  state (dotted). The dotted vertical lines at $z=0$
  and $z=d$ denote the bottom and top of the convectively
  unstable layer, respectively. 
  The last two panels show the turbulent rms-velocity and the
  energy of the toroidal mean magnetic field in the
  saturated phase.}
\label{fig:strat}
\end{figure}

\subsection{Nondimensional units and parameters}

Dimensionless quantities are obtained by setting
\begin{eqnarray}
d = g = \rho_0 = c_{\rm P} = \mu_0 = 1\;,
\end{eqnarray}
where $\rho_0$ is the initial density at $z_2$. The units of length,
time, velocity, density, entropy, and magnetic field are
\begin{eqnarray}
&& [x] = d\;,\;\; [t] = \sqrt{d/g}\;,\;\; [U]=\sqrt{dg}\;,\;\;
  [\rho]=\rho_0\;,\;\; \nonumber \\ && [s]=c_{\rm P}\;,\;\;
  [B]=\sqrt{dg\rho 
_0\mu_0}\;. 
\end{eqnarray}
We define the fluid and magnetic Prandtl numbers and the Rayleigh
number as
\begin{eqnarray}
\Pra=\frac{\nu}{\chi_0}\;,\;\; \Pm=\frac{\nu}{\eta}\;,\;\;
\Ra=\frac{gd^4}{\nu \chi_0} \bigg(-\frac{1}{c_{\rm P}}\frac{{\rm
    d}s}{{\rm d}z 
} \bigg)_0\;,
\end{eqnarray}
where $\chi_0 = K/(\rho_{\rm m} c_{\rm P})$ is the thermal
diffusivity, and $\rho_{\rm m}$ is the density in the middle of
the unstable layer. 
For the magnetic diffusivity we consider a $z$-dependent profile
which gives an order of magnitude smaller value in the radiative
layer than in the convection zone through a smooth transition. Thus
the magnetic Prandtl number in the stable layer is $\Pm=15$ and in the
convection zone $\Pm=1.5$.

The entropy gradient, measured in the middle of
the convectively unstable layer in the initial non-convecting 
hydrostatic state,
is given by
\begin{eqnarray}
\bigg(-\frac{1}{c_{\rm P}}\frac{{\rm d}s}{{\rm d}z}\bigg)_0 =
\frac{\nabla-\nabla_{\rm ad}}{H_{\rm P}}\;,\label{equ:super}   
\end{eqnarray}
where $\nabla-\nabla_{\rm ad}$ is the superadiabatic temperature
gradient with $\nabla_{\rm ad} = 1-1/\gamma$, $\nabla = (\pd \ln T/\pd
\ln p)_{z_{\rm m}}$, where $z_{\rm m}=\onehalf(z_3+z_2)$. The amount of 
stratification is determined by the parameter $\xi_0 =(\gamma-1) 
c_{\rm V}T_4/(gd)$, which is the pressure
scale height at the top of the domain normalized by the depth of the
unstable layer. We use in all cases $\xi_0 =0.12$, which results in a
density contrast of about 120. We define the fluid and magnetic
Reynolds numbers via
\begin{eqnarray}
{\Rey} = \frac{\urms}{\nu \kef}, \quad
{\Rm} = \frac{\urms}{\eta \kef}=\Pm\Rey,
\end{eqnarray}
where $\urms$ is the rms value of the velocity fluctuations and $\kef
= 2\pi/d$ is assumed as a reasonable estimate for the 
wavenumber of the energy-carrying eddies. Our definitions of the
Reynolds numbers are smaller than the usually adopted ones by a factor
of $2\pi$. The amount of shear is quantified by
\begin{eqnarray}
{\rm Sh} = \frac{U_0/d_{\rm s}}{\urms \kef},\label{equ:Sh}
\end{eqnarray}
where $U_0$ is the amplitude and $d_{\rm s}$ the width of the imposed
shear profile (see below). 
The equipartition magnetic field is defined by
\begin{equation}
\Beq \equiv
(\langle\mu_0\rho\bm{U}^2\rangle)^{1/2}_{z_{\rm ref}},\label{equ:Beq}   
\end{equation}
where the angular brackets denote horizontal average. For a better
comparison between the magnetic and kinetic energies, we evaluate
$\Beq$ at the center of the shear layer, $z=z_{\rm ref}$.

The simulations were performed with the {\sc Pencil Code}%
\footnote{\texttt{http://pencil-code.googlecode.com/}},
which uses sixth-order explicit finite differences in space and third 
order accurate time stepping method.

\subsection{Boundary conditions}

\begin{table*}[t]
\centering
\caption[]{Summary of the runs.}
\label{tab:1}
\vspace{-0.5cm}
$$
\begin{array}{p{0.035\linewidth}cccccccccccc}
  \hline
  \noalign{\smallskip}
  Run &  U_0/(dg)^{1/2} &  d_z & z_{\rm ref}  & \Sh & \Rey & \Rm &
  $Ma$ &  \lambda [10^{-2}]  
  & \Beq & \tilde{B}_{\rm rms} & \tilde{B_y} & \mean{B}/\mean{B}_{\rm T}
  \\ \hline 
  S00 &  0.00 &  0.05 & -0.05 & 0.0 & 8.9 & 14.7 & 0.028 & -2.16 &  0.06 &
  -  & - & -\\ 
  S01 &  0.06 &  0.05 & -0.05 & 3.7 & 8.2 & 13.6 & 0.026 & -1.29 & 0.05 &
  -  & - & -\\ 
  S02 &  0.09 &  0.05 & -0.05 & 6.8 &13.4 & 22.3 &0.042& -0.34 &  0.21 &
  -  & - &- \\ 
  S03 &  0.12 &  0.05 & -0.05 & 7.3 &15.3 & 25.5 &0.053 & 0.03 & 0.24 &
  -  & - & -\\
  S04 &  0.15 &  0.05 & -0.05 & 7.6 & 19.9 & 33.2 &0.062 & 1.72 & 
  0.40 & 0.31 & 5.32 &  0.34\\   
  \hline
  D01 &  0.15 &  0.05 & -0.05 & 7.6 & 19.9 & 33.2 &0.062 &   1.72 & 
  0.40 & 0.31 & 5.32 & 0.34\\
  D02 &  0.15 &  0.05 & -0.09 & 7.7 & 19.7 & 32.8 &0.062 & 1.72 &
  0.39 & 0.40  & 4.92 & 0.36 \\
  D03 &  0.15 &  0.05 & -0.13 & 7.6 & 19.9 & 33.1 &0.062 & 1.36 &
  0.39 & 0.42  &  4.83 &0.36\\
  D04 &  0.15 &  0.05 & -0.17 & 7.3 & 20.1 & 34.9 &0.066 & 1.37 &
  0.39 & 0.56  & 6.15 & 0.37\\
  \hline
  T01 &  0.15 &  0.05 & -0.05 & 7.6 & 19.9 & 33.2 &0.062 &  1.72 & 
  0.40 & 0.31 & 5.32 & 0.34\\
  T02 &  0.15 &  0.07 & -0.05 & 6.0 & 18.1 & 30.1 & 0.057 & 1.00 &
  0.33 & 0.34  & 4.79 & 0.36\\
  T03 &  0.15 &  0.09 & -0.05 & 4.9 & 17.2 & 28.7 & 0.054 & 0.91 &
  0.30 & 0.31  & 5.40 & 0.37\\
  T04 &  0.15 &  0.11 & -0.05 & 4.4 & 15.8 & 26.3 & 0.049 & 0.88 &
  0.26 & 0.44  & 6.15 &  0.41\\
  T04$_0$ &  0.15 &  0.11 & -0.05 & 4.1 & 16.6 &  20.8 & 0.052 & -4.3 &
  0.28 & -  & - & - \\
  T04$\tau$ &  0.15 &  0.11 & -0.05 & 3.8 & 18.1 & 30.2 & 0.056 & 1.18 &
  0.36 & 0.40  & 3.75 & 0.39 \\
  T05 &  0.15 &  0.15 & -0.05 & 3.4 & 15.0 & 24.9 & 0.045 & 0.39 &
  0.24 & -  & - & - \\ 
  T06 &  0.15 &  0.20 & -0.05 & 2.9 & 13.1 & 22.0 & 0.041 & -0.04 &
  0.19 & -  & - & - \\ 
\hline
\end{array}
$$
\tablefoot{
  Here $\mbox{Ma}=\urms/(gd)^{1/2}$,
  $\lambda$ is the growth rate in units of $\tau_{\rm turn}$,
  $\tilde{B}_{\rm rms} \equiv \brms/B_{\rm eq}$, $\tilde{B}_y \equiv
  {\rm max}(|B_y|)/B_{\rm eq}$, where the equipartition
  magnetic field is evaluated at $z=z_{\rm ref}$ (see
  Eq. \ref{equ:Beq}) and is given in units of $\sqrt{dg \rho_0 \mu_0}$.
  Finally $\mean{B}/\mean{B}_{\rm T} = \sqrt{\brac{{\bm B}}_y^2 /\brac{\bm B^2_y}} $
  corresponds to fraction of mean magnetic field compared with the
  total field.  
  The numbers are given for the saturated state of the dynamo. In all 
  these simulations the resolution is   $128^3$ grid points, 
  $\Pra=20$ and $\Ra\simeq8.3\times10^6$.
}
\end{table*}

In the horizontal $x$ and $y$ directions we use periodic boundary
conditions and at the vertical ($z$) boundaries we use stress-free
boundary conditions for the velocity,
\begin{equation}
U_{x,z} = U_{y,z} = U_z = 0.
\end{equation}
For the magnetic field vertical field condition is used on the
upper boundary whereas perfect conductor conditions are used at the
lower boundary, i.e.\,
\begin{eqnarray}
B_x = B_y &=& 0, \;\; (z=z_4) \\
B_{x,z} = B_{y,z} = B_z &=& 0, \;\; (z=z_1)
\end{eqnarray}
respectively. The upper boundary thus allows magnetic helicity flux
whereas at the lower boundary does not. This is likely to be
representative of the situation in a real star where magnetic helicity
can escape via the surface but does not penetrate into the core.

\subsection{Shear at the base of the convection zone}
\label{sec:shear}
In order to mimic the tachocline at the base of the solar convection
zone we introduce a shear profile
\begin{eqnarray}
\meanv{U}^{(0)}=\onehalf U_0 \tanh \left(\frac{z-z_{\rm ref}}{d_{\rm s}}\right)\hat{\bm{e}}_y,
\end{eqnarray}
where $z_{\rm ref}$ is the reference position of the shear
layer. Given the uncertainties of the radial position and width of the
tachocline we perform parameter studies where $z_{\rm ref}$ and
$d_{\rm s}$ are varied. Furthermore, it is of general interest to
study how dynamo excitation is depends on varying $U_0$ and the ratio
of $U_0/d_{\rm s}$.

\section{Results}
\label{sec:results}

With the purpose of addressing the questions raised in the introduction we 
perform a series of simulations with the model described above where
some properties of the shear layer are varied. We first study
the hydrodynamic properties of the system and the conditions for
dynamo excitation. The results of this parameter study are
summarized in Table~\ref{tab:1}. Then we study the topological and 
buoyant properties of the magnetic fields generated in some
characteristic runs and compare them with models with different aspect
ratio and higher resolution (see Table~\ref{tab:2}). We finalize
this section with the study of the magnetic feedback on the plasma
motion and the computation of the turbulent coefficients that govern
the evolution of large-scale magnetic fields in our simulations.

\subsection{Hydrodynamic instabilities}
In our simulation setup, two kinds of hydrodynamical instabilities may
develop, namely the Kelvin-Helmholtz (KH) instability due to the
imposed shear and stratification, and the convective instability due
to the superadiabatic stratification in the middle layer. From
hydrodynamical runs, we find that the convective instability develops
early, at $t \urms \kef \approx 40$, whereas the KH-instability
develops at $t \urms \kef \approx 100$ in runs with the
strongest shear. After a few hundred time
units the velocity reaches a statistically steady state (constant
rms-velocity),
and the toroidal velocity achieves the desired shear profile.
A thermally relaxed state (constant thermal energy), 
however, is reached only after a few thousand time units. This time
depends on the radiative conductivity ($K$) but also on the
imposed shear, which produces viscous heating that
modifies the thermal stratification of the system as it
may be seen in the top panels of Fig.~\ref{fig:strat}. 
The final velocity profile which includes shear, convection, and
the KH instability, does not allow the possibility of including 
rotation in the model. If it is done, the system develops
mean field motions in the horizontal direction which are undesirable
in the present study.

\begin{table*}[t]
\centering
\caption[]{Summary of runs with different aspect ratio (upper two
  rows) and higher resolution (lower rows).}
\label{tab:2}
\vspace{-0.5cm}
$$
\begin{array}{p{0.035\linewidth}ccccccccccccccc}
  \hline
  \noalign{\smallskip}
  Run & (L_x,L_y,L_z) & U_0/(dg)^{1/2} &  d_z & z_{\rm ref} & $Pr$  & \Sh & \Rey
  & \Rm & \Ma_{\rm k} & \Ma_{\rm s} &  \lambda [10^{-2}] & \Beq &
  \tilde{B}_{\rm rms} & \tilde{B_y}  & \mean{B}/\mean{B}_{\rm T}
  \\ \hline 
  AR01 & (8,4,2)d  &  0.15 &  0.11 & -0.05 & 20 & 4.4 & 15.8 & 26.3 &
  0.050 & 0.050 & 0.89 &  0.25 & 0.50  & 6.08  &0.40 \\ 
  AR02 & (4,8,2)d &  0.15 &  0.11 & -0.05 & 20 & 3.8 & 18.4 & 30.6 &
  0.058 & 0.058 & 0.30 &  0.32 & 0.22  & 4.73 & 0.29 \\ 
  \hline
  D03b & (4,4,2)d &  0.15 &  0.05 & -0.10 & 12 & 7.9 & 31.8 & 47.7  &
  0.062 & 0.060 & 2.57  &  0.43 & 0.31  & 4.16 & 0.31 \\
  T04a & (4,4,2)d &  0.15 &  0.1 & -0.05 & 20 & 5.1 & 25.0 & 37.6  &
  0.047 & 0.047 & 2.23 & 0.26 & 0.41  & 5.06 & 0.36 \\
  T04b & (4,4,2)d &  0.15 &  0.1 & -0.05 & 12 & 4.9 & 25.6 & 38.5  &
  0.052 & 0.048 & 2.19  & 0.27 & 0.43  & 4.80 & 0.37 \\
  T04c & (4,4,2)d &  0.15 &  0.1 & -0.05 & 3 & 4.4 & 28.6 & 43.0  &
  0.056 & 0.054 & 1.21  & 0.30 & 0.35  & 4.13 & 0.36 \\\hline
\end{array}
$$
\tablefoot{Most of the quantities here are defined in 
  Table~\ref{tab:1}.  We have added here the scales of the domain, the
  Prandtl number and the Mach number at the kinematic ($\Ma_{\rm k}$) and
  saturated ($\Ma_{\rm s}$) stages. Models with different aspect ratio have
  the same spatial resolution than models in
  Table~\ref{tab:1}. In Runs~D03b and T04a-c the resolution is 256$^3$
  grid points.  
}
\end{table*}

\subsection{Dynamo excitation}
In order to study the dynamo excitation we follow the evolution of an
initial random seed magnetic field of the order of $10^{-5}\Beq$.
Since there is no rotation and because the vertical component of the
large-scale  vorticity $\mean{W}_i=\epsilon_{ijk}\mean{U}_{j,k}\approx
0$, the average kinetic helicity and mean-field $\alpha$-effect are
expected to vanish \citep[e.g.][]{KR80}. However, it is still possible
to excite a large-scale dynamo due to non-helical turbulence and shear
\citep{bran05,yousef_etal_08b,yousef_etal_08a,BRRK08}.
In this part of this study we vary three parameters:
the differential rotation amplitude, $U_0$ (Runs~S00--S04), the
location of the shear layer in the convectively stable region, $z_{\rm
  ref}$  (Runs~D02--D04), and the thickness of the shear region, $d_z$
(Runs~T02--T04).  
 
\begin{figure}[t]
\centering
\includegraphics[width=.99\columnwidth]{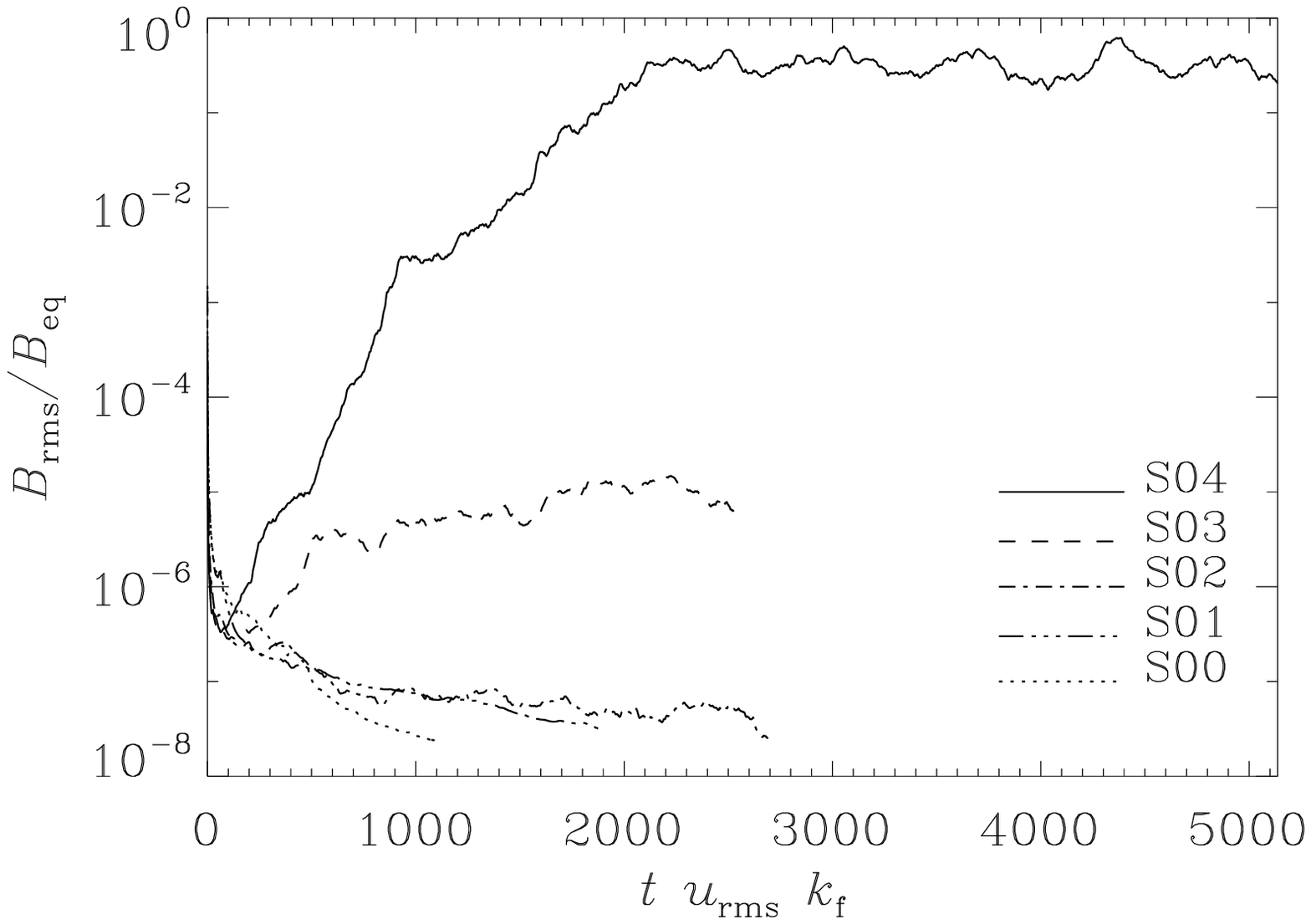}
\includegraphics[width=.99\columnwidth]{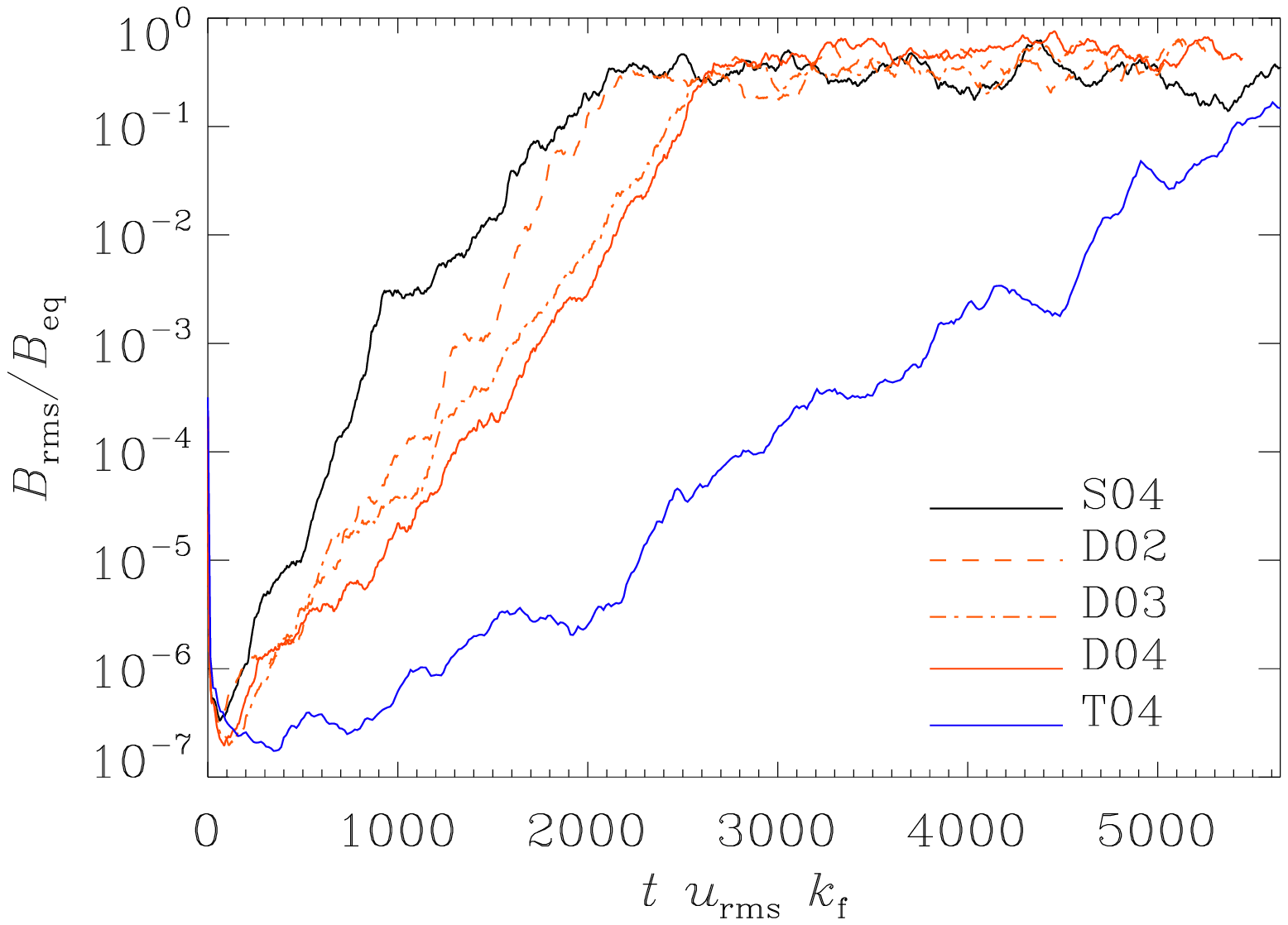}
\includegraphics[width=.99\columnwidth]{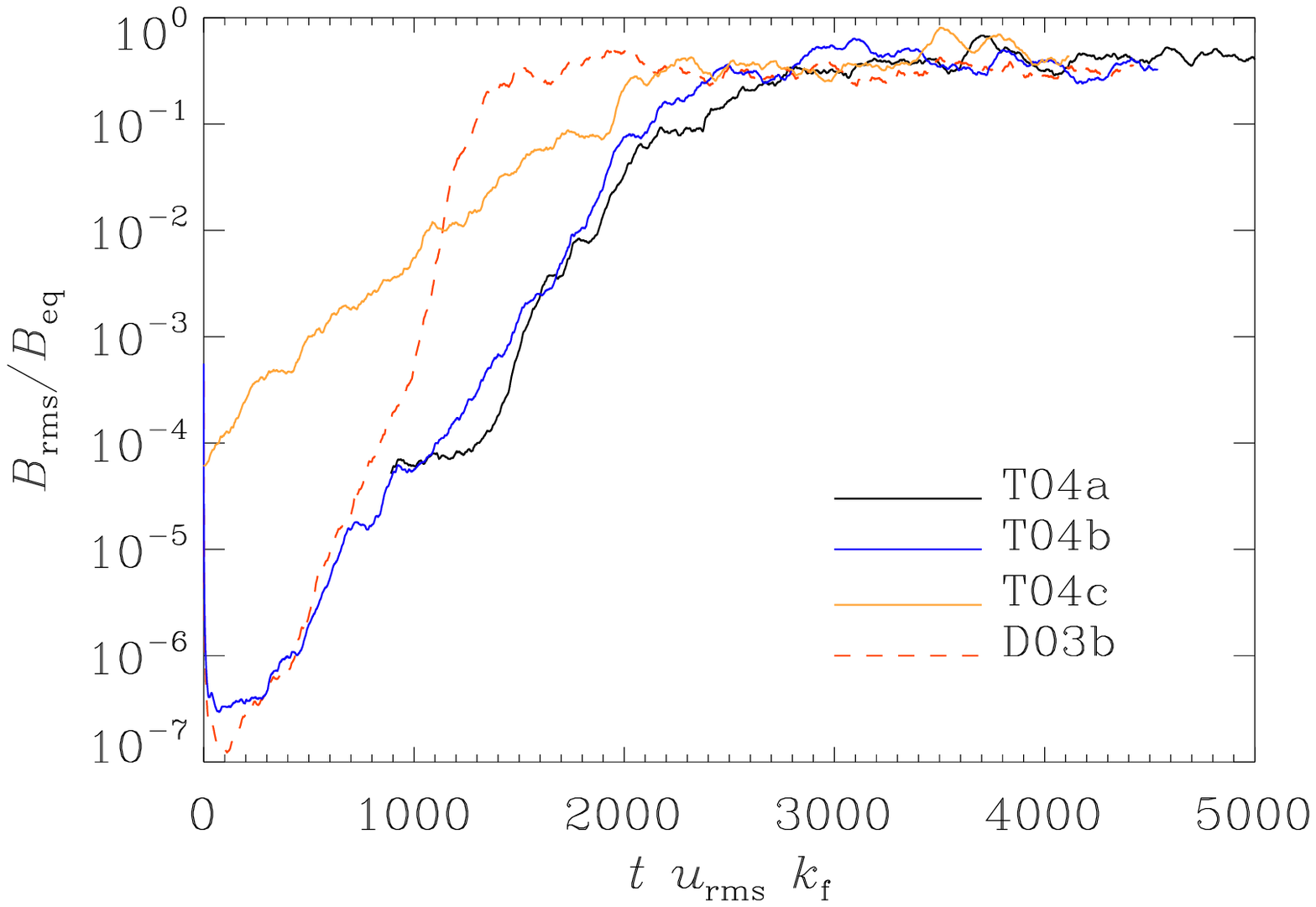}
\caption{Time evolution of the rms value of the total magnetic field
  normalized with the equipartition value at the center of the shear
  layer. Different lines/colors correspond to different runs as
  indicated in the legends in each panel.}
\label{fig:b-t}
\end{figure}

We note that no small-scale dynamo is excited in the non-shearing
case. With the runs in Set~S we find that the critical value of
shear above which a dynamo is excited is $\Sh\approx7$ (see the top
panel of Fig.~\ref{fig:b-t}). 
The value of $\Sh$ is roughly three times larger than that
computed with typical values of the rms-velocity and length scales in 
the solar tachocline \footnote{We have considered
  $\Delta\Omega=33$nHz, $r_{\rm tac}=0.7R_{\odot}$,
  $d_{\rm tac}=0.05R_{\odot}$ \citep[e.g.][and references
    therein]{CD+Thomp_07}, $\urms=10^4$ cm s$^{-1}$, and $k_{\rm f} = 
  2\pi/l$, with $l=10^{10}$ cm \citep[see Table 1 of][]{Bran+Sub_05},
  which result in $\Sh_{\rm tac}\approx 2$.}.

In Set~D, we move the
shear layer deeper down into the stable layer, where the
turbulent diffusivity is expected to be less, thus allowing a longer 
storage of the magnetic field\footnote{Note that Run~S04 is the same as
  Runs~D01 and T01.}.
From the hydrodynamical point of view, these simulations have a
convection zone that, due to viscous heating, extends into the 
initially stable layer in the saturated state
as it can be seen in the vertical profile of entropy
corresponding to Run~D04 (red line) in Fig.~\ref{fig:strat}). 
The three bottom panels of the  same figure (see also
Table~\ref{tab:1}) show (from top to bottom) the
vertical profiles of the mean toroidal velocity ($\mean{U}_y$),
the turbulent rms-velocity and $\mean{B}_y$ in the relaxed
state. 
Through this paper we consider averages taken first over the $y$
direction and then the averaging over the other directions is
performed.
Note that the $\urms$-profiles have been computed neglecting
the mean flow. 
The curves of the turbulent velocity in the models that include the
shear layer exhibit a bump where the velocity gradient is located.
This feature does not appear in non-shearing models (see dotted line),
suggesting that turbulent motions are developing in this regions. 
The bump is more pronunced in models with larger $\Sh$ which indicates
that this turbulent motions are probably due to the KH instability. 
This difference in $\urms$ reflects in the value of the
equipartition magnetic field (Eq.~\ref{equ:Beq}) which differs from
model to model.

We find that the amplitude of the magnetic energy increases for deeper
tachoclines. This is an expected result since below the convection
zone the magnetic diffusivity has smaller values which allows a larger
storage. The vertical distribution of the toroidal component of the
magnetic energy ($\mean{B}_y^2$) peaks roughly at the center of the
shear layer. The tail of this curve towards the convection zone may
hint how buoyant the magnetic field is on each
model. We expect that the larger the magnetic field (Run~D04) the 
more magnetic flux becomes buoyant and rises into the 
upper layers.   

The time evolution of the magnetic field of the runs in Set~D
are shown in the middle panel of Fig.~\ref{fig:b-t}. We find
that the growth rate of the magnetic field, $\lambda={\rm d}\ln
\brms/{\rm d}t$, decreases slightly as the shear layer is moved
deeper. This is not very clear in Table~\ref{tab:1} since the error
of this quantity is of the same order as the 
difference between different runs. 
However, it is clear from both, the figure and the table, that the
volume averaged magnetic field 
increases with the tachocline depth.

In the third group of simulations (Runs~T01--T04), we increase the
width of the shear layer gradually from $d_z=0.05d$ in Run~T01 
to $d_z=0.11d$ in Run~T04. This change implies lower values of the
shear parameter $\Sh$ (Eq.~\ref{equ:Sh}), but also larger fraction of the
tachocline in the turbulent and stable layers.  The results, presented
in Table~\ref{tab:1} and depicted in Figs.~\ref{fig:strat} and
\ref{fig:b-t} (see legends), indicate that smaller values of the
shear, $\Sh\approx3$,
are still able to excite the dynamo at the price of a lower growth
rate.  We notice that the growth rate
depends on the magnetic Reynolds number (compare Runs~T04 and T04b).
The fact that a smaller shear generates a larger magnetic field is 
a counter-intuitive result and does not agree with previous mean-field
studies on the thickness of the solar tachocline \citep{gue07a}.  
However,
as it will be explained below, 
the important fact is that
this configuration seems to be
favorable to a longer storage of magnetic field in the stably
stratified layer, with the advantage that
here the effects of the shear on the thermal properties of the fluid
are less important than in the previous sets of simulations.
With the settings of Run~T04 we find that the critical magnetic Reynolds 
number is between $20.8$ (Run~T04$_0$) and $26.3$ (Run~T04$_0$).

Based on the results above, we perform another series of 
simulations whose parameters and results are summarized in 
Table~\ref{tab:2}. Runs~AR01 and AR02 correspond to Run~T04 but with
aspect ratios $(L_x,L_y,L_z)=(8,4,2)d$ and $(4,8,2)d$, 
respectively. Runs~D03b, and T04a to T04c have
essentially the same configuration than Runs~D03 and T04 with $256^3$
grid points resolution. 
The values of $\nu$ and $K$ have been modified in order to obtain
different Reynolds and Prandtl numbers. 

The model with larger extent perpendicular to the direction of the
shear does not show differences with respect to the reference case. On
the other hand, the model with larger extent in the direction of the
shear results in a reduced growth rate and in smaller values of the
volume averaged magnetic field ($\tilde{B}_{\rm rms}$), the maximum
amplitude of the toroidal magnetic field and of the large-scale
field. The reason for these changes is the increase of the $\urms$
velocity, which diminishes the effective shear. 
\cite{yousef_etal_08b} have obtained that both the growth rate and
the scale where the maximum magnetic energy is concentrated, converge
to a value which is independent of the relevant length scale ($L_z$ in
their case). Such convergence analysis is 
computationally very expensive to be performed for our system. The
models with higher resolution (see Runs~D03b and T04b in 
Table~\ref{tab:2} and bottom panel of Fig.~\ref{fig:b-t}) show a larger
growth rate when compared with their corresponding low resolution
cases. These runs, however, saturate at sligthly smaller amplitudes of
$\tilde{B}_{\rm rms}$. 
Similar weak dependence on $\Rm$ has been reported from simulations
with horizontal shear \citep{kkb10a}.

We have also changed the $\Rm$ by considering a different input heat
flux (i.e.\ different $\Pr$). The results of Runs~T04a--T04c show that
$\lambda$ depends on $\Pr$. There is no considerable
difference between the rms magnetic field of these runs.
 
\subsection{Morphology of the dynamo generated magnetic field} 
\label{sec:buoyancy}
\begin{figure}[t]
\centering
\includegraphics[width=.99\columnwidth]{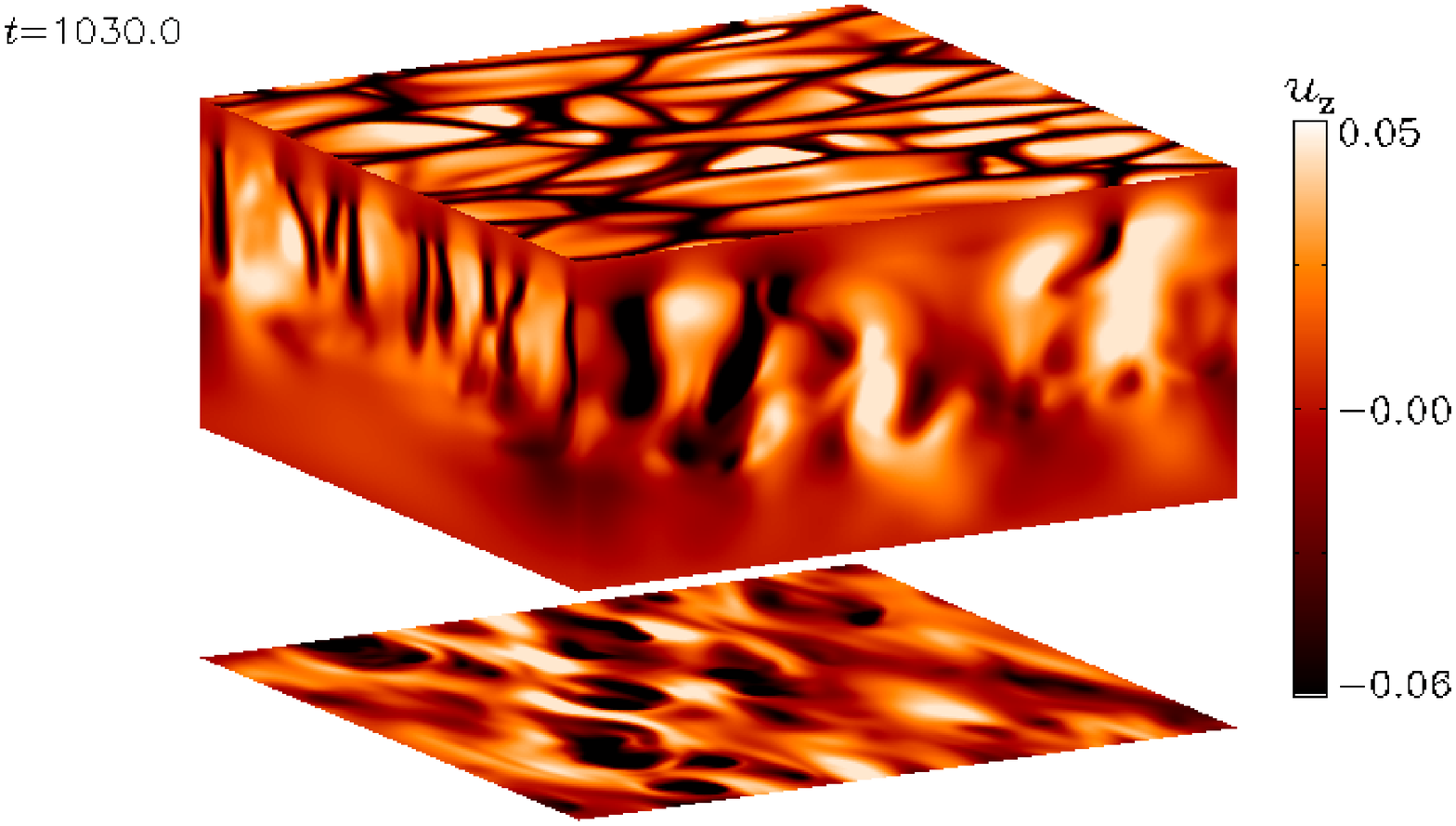}\\
\includegraphics[width=.99\columnwidth]{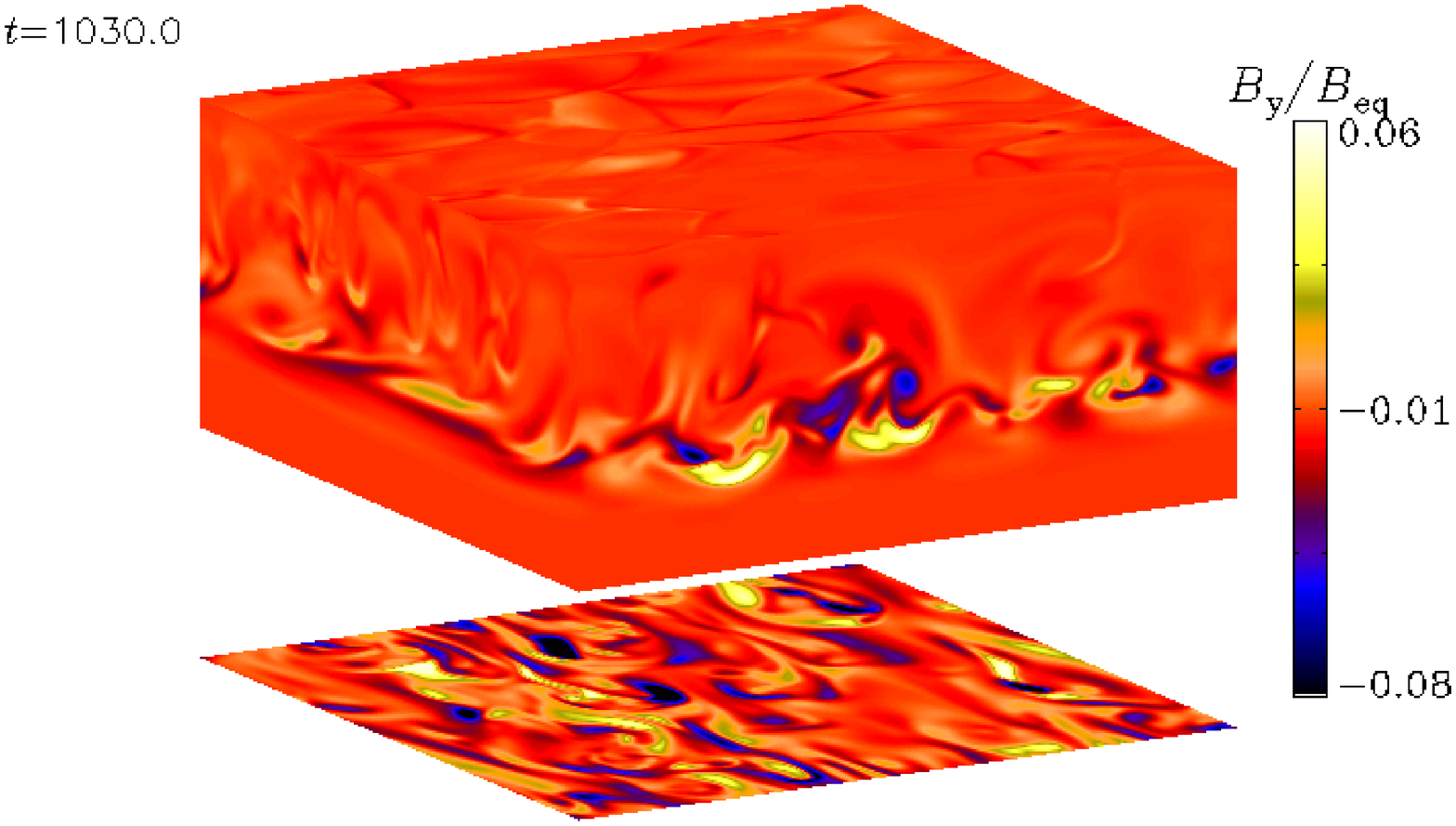}
\caption{Snapshots of the vertical velocity (top row) and the
  toroidal magnetic field ($B_y$) in different planes of the domain,
  for Runs~T04c (from left to right) in the kinematic
  phase. The vertical slices correspond to
  $yz$ (left side of the box) and $xz$ (front of the box) planes. The top
  plane corresponds to the horizontal boundary between the convective
  and the cooling layers at $z=d$ and the plane shown below the box
  corresponds to the base of the unstable layer ($z=0$). }
\label{fig:byuz_kin}
\end{figure}

In the initial stages of evolution, the magnetic field
is dominated by small scales and although it is 
possible to distinguish the stretching effects of the shear, the
structures formed are small compared with the size of the box. 
In Fig.~\ref{fig:byuz_kin} we present snapshots of the Run~T04c in the
kinematic phase.
   
\begin{figure}[t]
\centering
\includegraphics[width=.99\columnwidth]{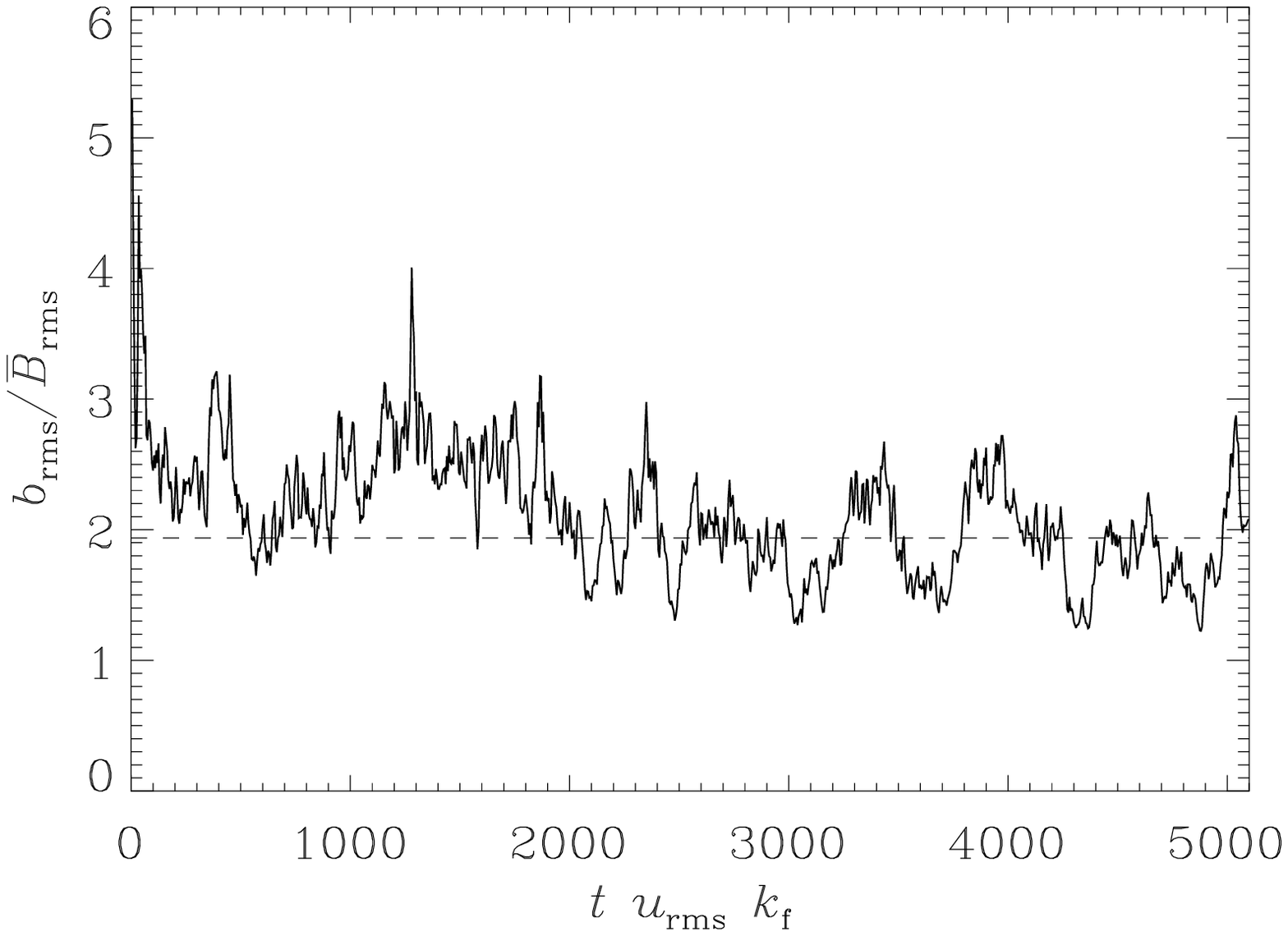}
\includegraphics[width=.99\columnwidth]{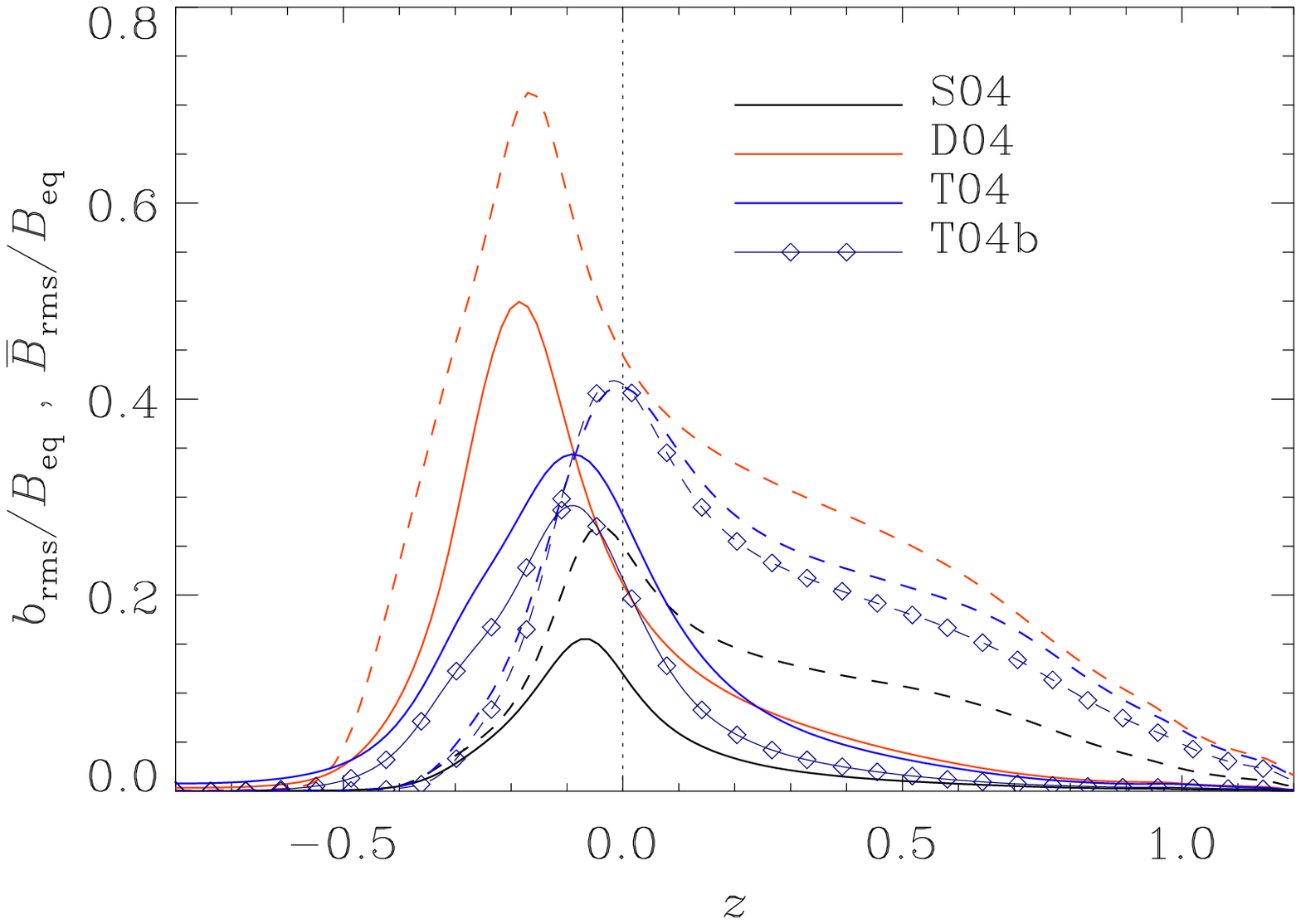}
\caption{Upper panel: ratio between small scale, $b_{\rm rms}$, 
and mean, $\mean{B}_{\rm rms}$, magnetic fields for
Run~S04. The dashed line corresponds to the mean value of this ratio
in the saturated phase.  Lower panel: radial distribution of the mean
(thick lines) and fluctuating (thin lines) magnetic field for the
Runs~S04, D04 and T04 in the saturated phase.}
\label{fig:fvsm}
\end{figure}
Mean magnetic fields take longer to develop in all 
simulations but in the saturated state they correspond to a
considerable fraction of the total field. This can be seen in the
upper panel of Fig.~\ref{fig:fvsm}, where the ratio $b_{\rm
  rms}/\mean{B}_{\rm rms}$ for Run~S04 is shown as a function of
time.  
It is important to notice that mean values refer here to
averages over the $y$ direction, i.e.\ $\mean{B}_{\rm rms}=\sqrt{\brac{{\bm 
      B}}_y^2}$, and  $b_{\rm rms}=\sqrt{\brac{{\bm B}^2}_y - \brac{{\bm
      B}}_y^2}$. We do not perform average over the $x$
direction because the toroidal magnetic field varies also in $x$ (see
below). The average over $x$ and time is performed
after the mean and fluctuation values are computed. 

In this figure, the dashed line corresponds to the average
value of this ratio ($1.9$) in the saturated phase of the dynamo,
indicating that the mean magnetic field is roughly a third 
of the total magnetic field.  
For the sake of clarity only a single run is shown in this panel, but
similar results are obtained for all simulations in Sets~D and T.

In the bottom panel of Fig.~\ref{fig:fvsm}, the vertical profiles of the
mean (solid lines) and fluctuating (dashed lines) magnetic fields for
four representative runs are shown. We find that
the mean magnetic field is mainly located in the shear region, whereas the
turbulent field, on the other hand, is more spread out inside the convection zone. It is noteworthy that for a thick tachocline the mean field is 
almost comparable with the fluctuating one, covering a fraction of the 
stable layer where the fluctuations are weak (see continuous
blue line). 
Simulations with higher resolution (see blue lines with diamond
symbols in Fig.~\ref{fig:fvsm}) exhibit fluctuating magnetic field
with vertical distribution and amplitude almost identical to the
lower resolution case. The vertical profile of the mean
field is roughly the same as in the lower resolution case but of
smaller amplitude.

\begin{figure*}[t]
\centering
\includegraphics[width=.67\columnwidth]{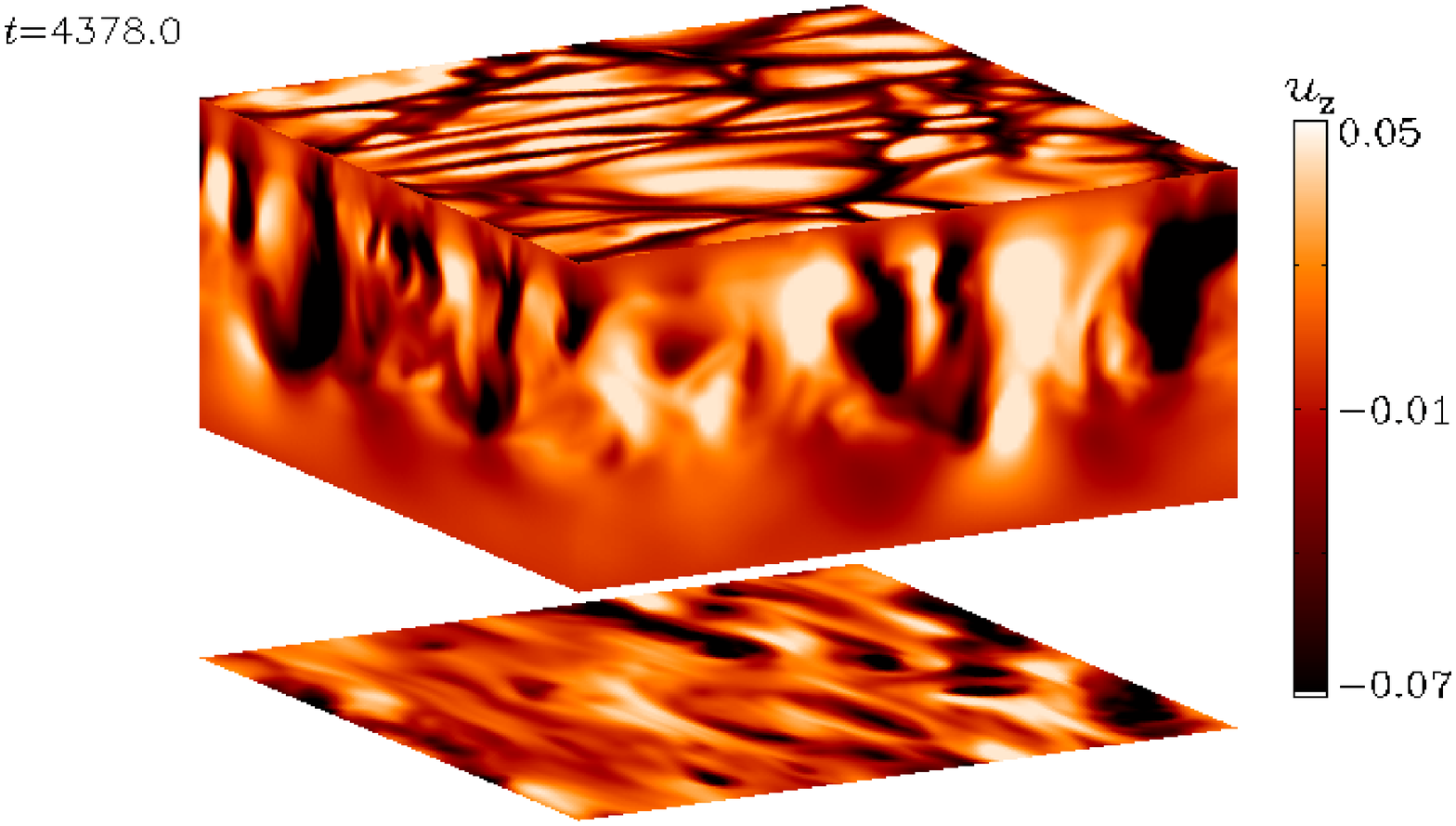}
\includegraphics[width=.67\columnwidth]{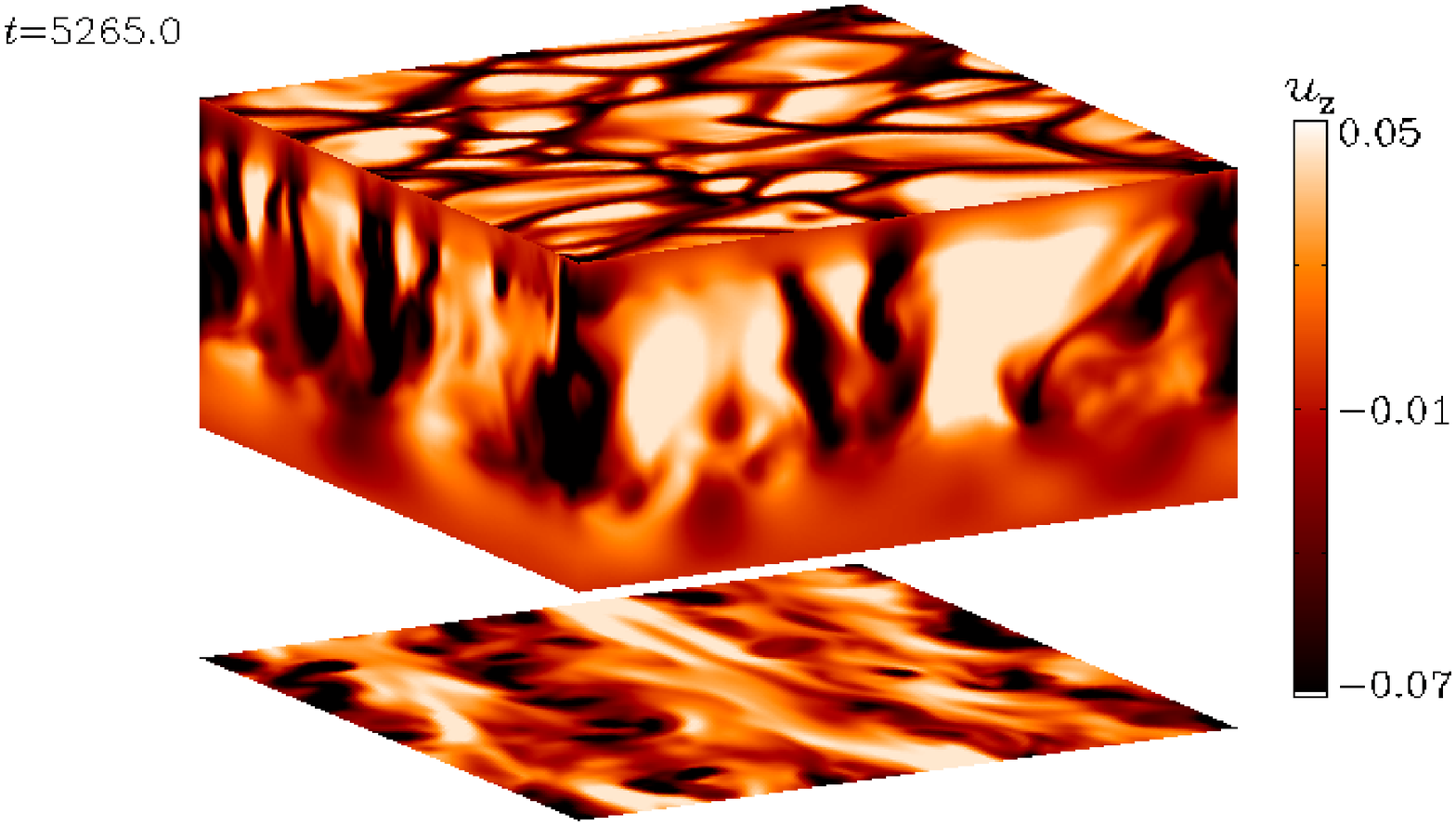}
\includegraphics[width=.67\columnwidth]{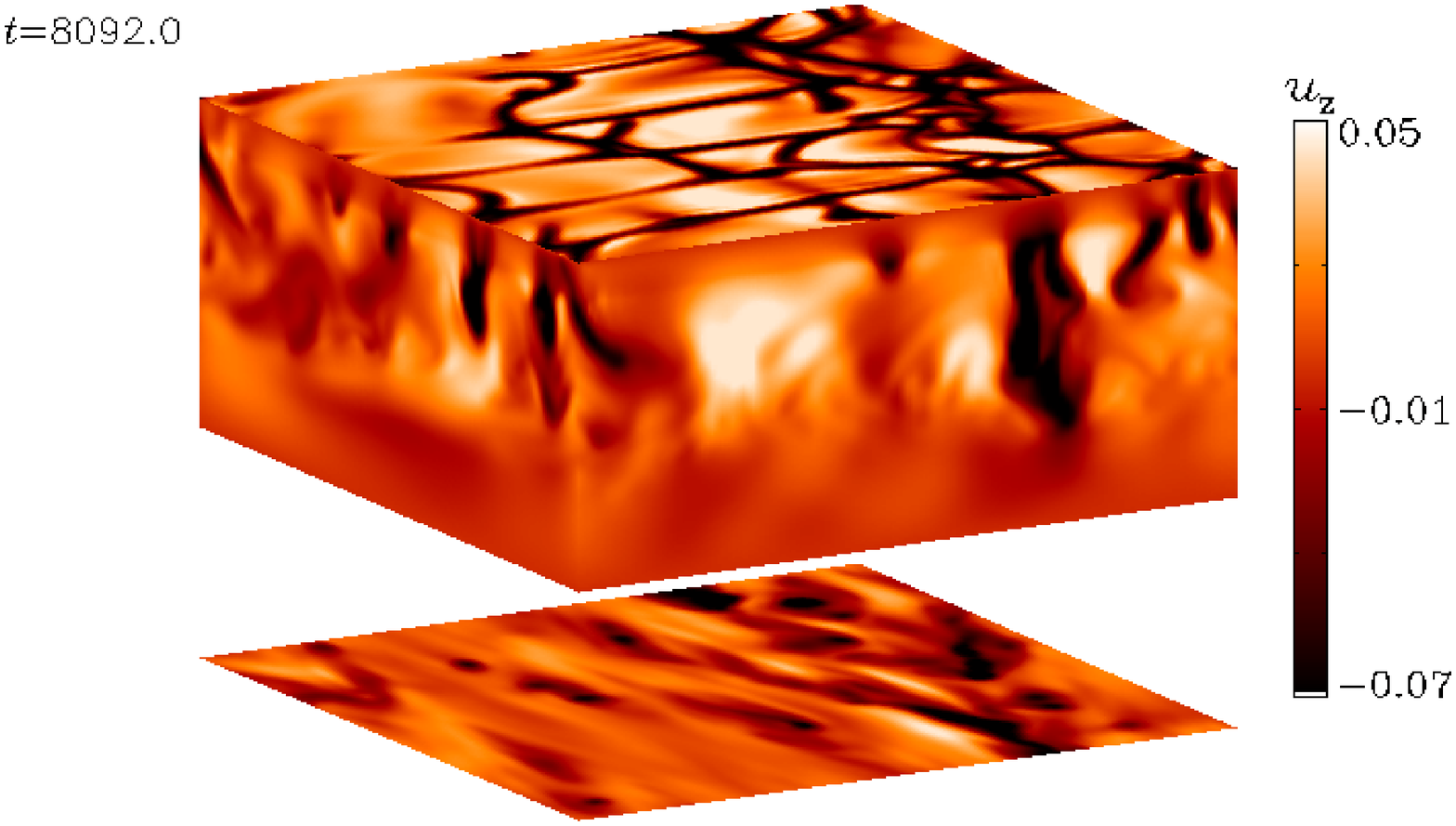}\\
\includegraphics[width=.67\columnwidth]{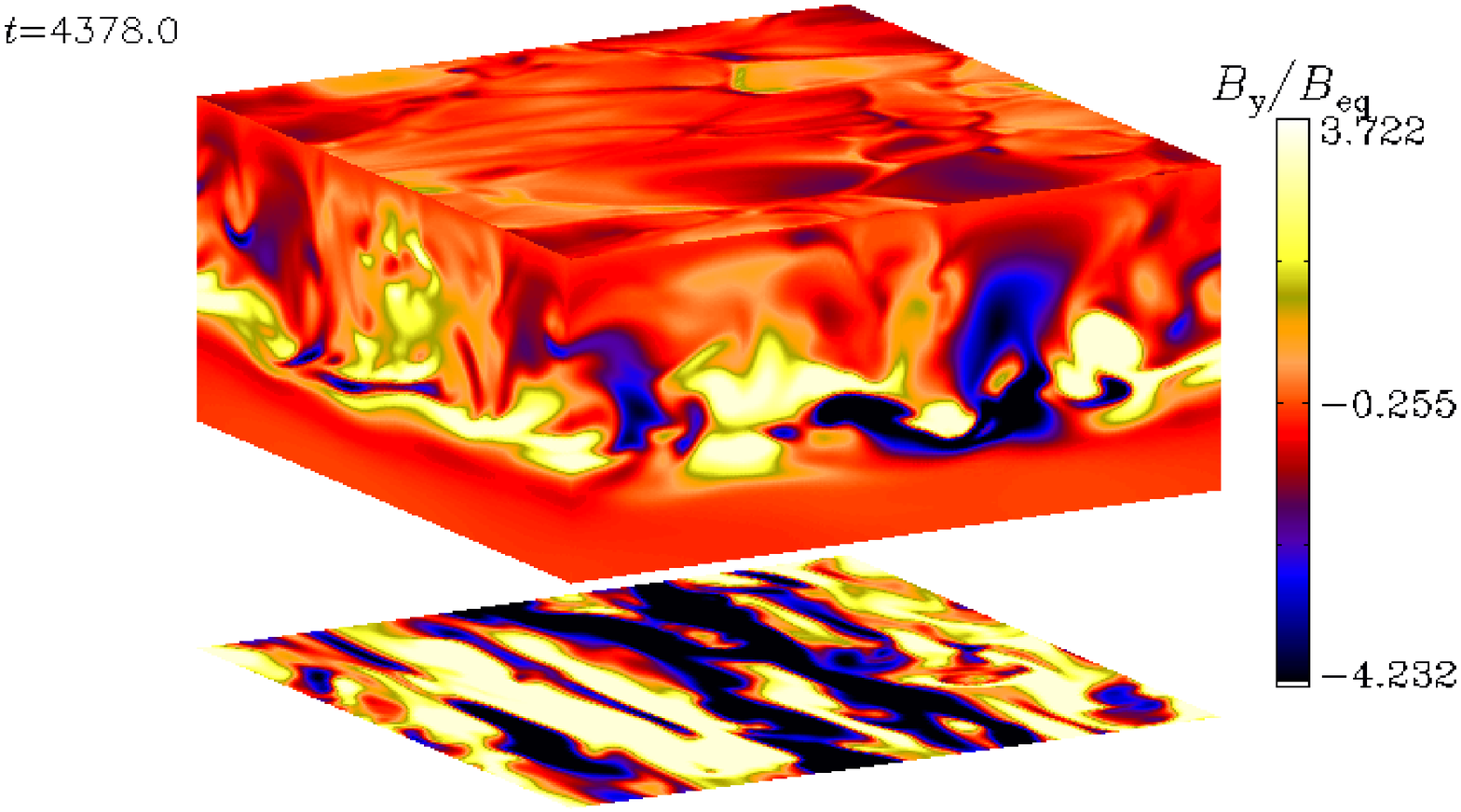}
\includegraphics[width=.67\columnwidth]{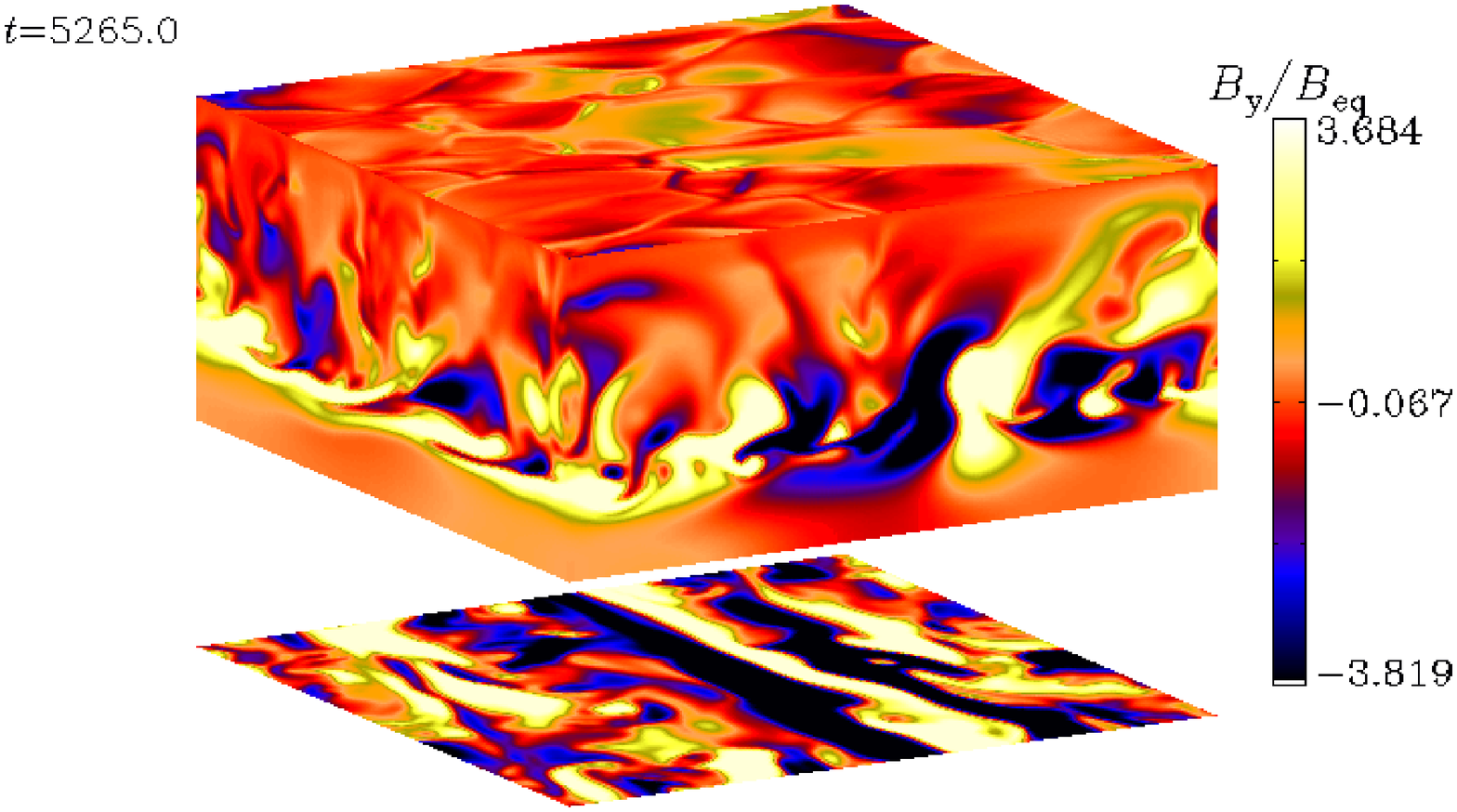}
\includegraphics[width=.67\columnwidth]{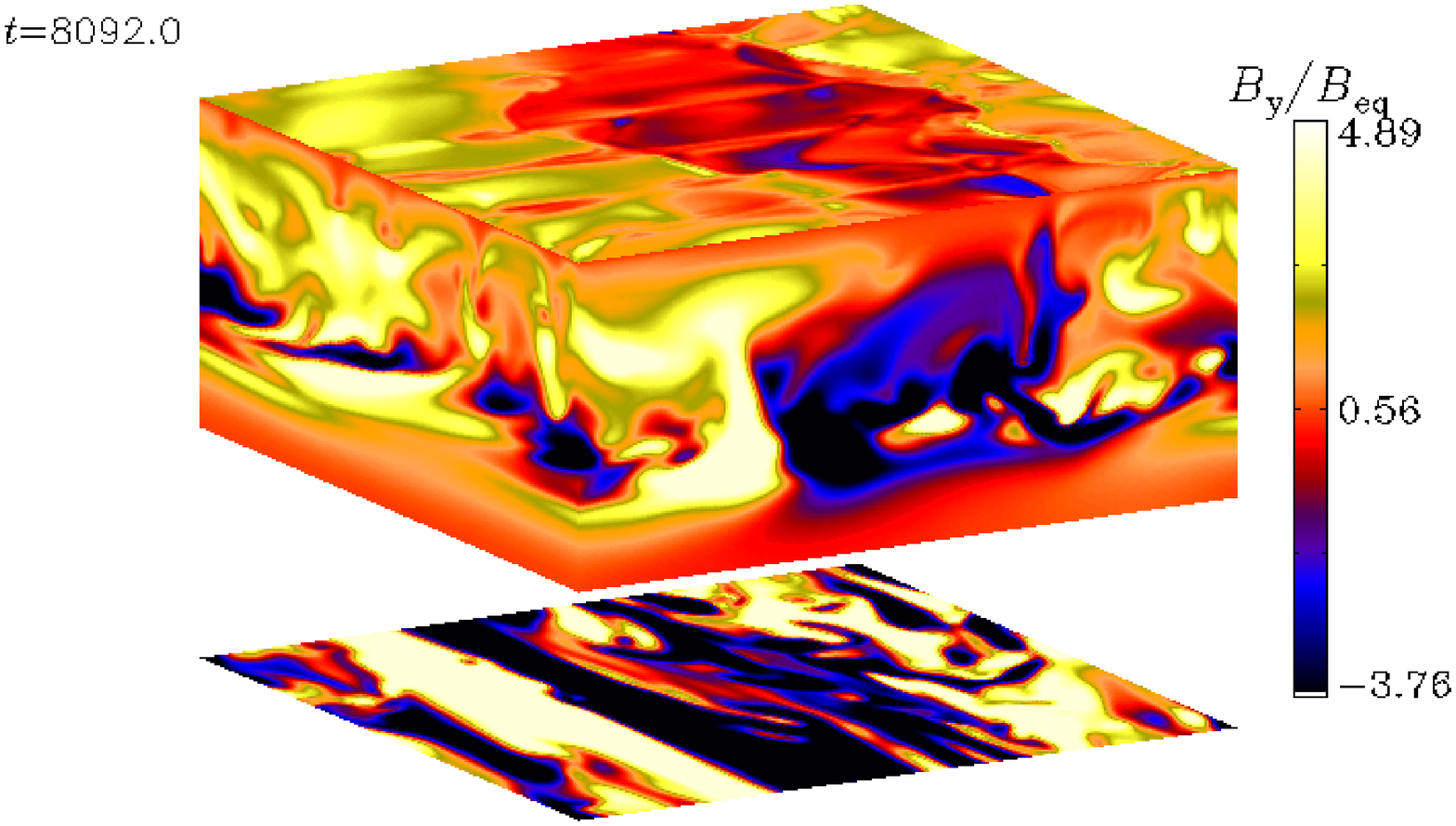}
\caption{
  Same as Fig.~\ref{fig:byuz_kin} but for Runs~S04, D04 and T04 (from left to right)  in the saturated phase. Movies corresponding to these runs may be found at: www.nordita.org/$\sim$guerrero/movies.}
\label{fig:byuz}
\end{figure*}

The structure of the magnetic fields depends strongly on the 
structure of convection. In
Fig.~\ref{fig:byuz} we show snapshots of vertical velocity ($U_z$) and
azimuthal magnetic field ($B_y$) for arbitrary times in the
saturated state for Runs~S04, D04, and T04 (from
left to right, respectively). 
From the bottom panels of this figure, it is possible to
distinguish that at the base of the convection zone, the 
toroidal field organizes in elongated structures of both polarities
which span all across the azimuthal direction if the
penetrative downflows are less intense. These stripes coexist
with more random magnetic fields in regions located where jet-like 
overshooting is able to reach the stable layer.  The size of the
magnetic structures is at least equal or larger than the scale of the
convective eddies. This is more 
evident for the streamwise direction where the field occupies
the full extent of the domain.
The strong toroidal fields are mainly confined in the shear region.
This can be seen in the bottom panels of Fig.~\ref{fig:byuz} and also in the
bottom panel of Fig. \ref{fig:strat}, where it is clear that the curve
corresponding to Run~T04 (blue line) has a broader profile than the
one corresponding to Run~S04  
(black line). The presence of the KH instability is
evident in the $yz$ plane where wave like structures are
observed. If the tachocline is thicker, the effects of the KH-instability
are weaker. Another advantage of thicker shear layers is that they do
not affect the thermodynamical properties of the fluid
as strongly as in Sets~S and T.
The disadvantage is that they produce broad and more diffuse magnetic fields which do not totally agree with the picture of a flux tube. This is discussed in more detail in the next section.

These results are in contradiction with those of \cite{Tobias+etal_08} who used a similar setup in the Boussinesq approximation. We believe that the lack of a large-scale dynamo in their simulations may be due either to the wider shear profile that they used or to the smaller amplitude of the shear parameter. However, a visual inspection of their figure 11 suggests that a large-scale dynamo could indeed exist in the horizontal plane. 

\subsection{Buoyancy}
\label{sec:buoyancy}
\begin{figure*}[t]
\centering
\includegraphics[width=2\columnwidth]{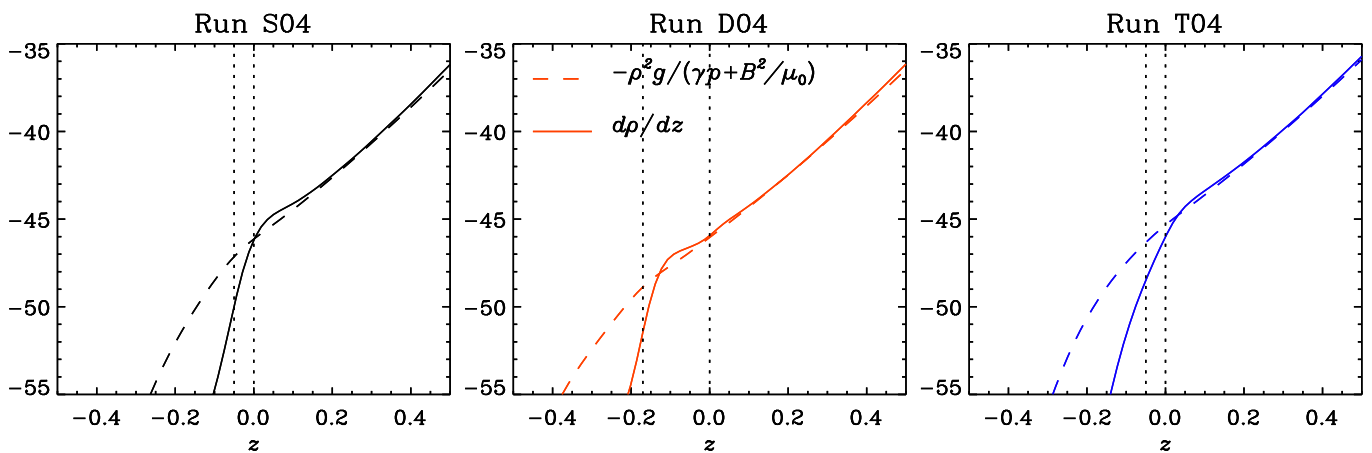}
\caption{Buoyancy instability condition, Eq.~(\ref{eq:binst}) evaluated
  for Runs~S04, D04 and T04. Continuous (dashed) lines correspond to
  the left (right) hand side of the equation, respectively. Dotted
  lines correspond to the interfaces between convectively stable and
  unstable layers, and to the center of the tachocline in
  each simulation.}
\label{fig:bi}
\end{figure*}

According to linear theory \citep{spruit+vanbal_82}, thin
magnetic flux tubes become buoyantly unstable when
the condition $\beta \delta > -1/\gamma$ is fulfilled. Here
$\beta=2\mu_0 p/B^2$ is the plasma beta, 
$\delta = \nabla - \nabla_{\rm ad}$ is the
superadiabaticity (see Eq. \ref{equ:super}), and $\gamma$ is the 
ratio of specific heats. This condition is satisfied
in the convection zone where $\delta > 0$. From this relation it
follows that below the convection zone, where $\delta < 0$, thin
magnetic flux tubes require much higher field strength to become
unstable. For a magnetic layer, the necessary and sufficient
condition for the development of two-dimensional interchange modes,
in which we are interested here, is given by \citep{Newcomb_61}:
\begin{equation}
\frac{d\rho}{dz} > \frac{-\rho^2 g}{\gamma p + B^2/\mu_0} ~.
\label{eq:binst}
\end{equation}
We find that for the cases with
thinnest shear layer, the instability region is above the center of
the shear layer where the toroidal magnetic field reaches its maximum
value (see Fig.~\ref{fig:bi}). This implies that a large fraction 
of the magnetic field
there becomes buoyantly unstable very quickly. For the
cases with a thick shear layer (Set~T), the magnetic field is
unstable only within the convection zone.
The magnetic field distribution of the Run~T04 in
Fig. \ref{fig:fvsm} (blue line), indicates
that a large fraction of the magnetic field is buoyantly
stable. This configuration may explain why in these cases larger
mean magnetic fields develop with smaller shear parameters  
than in sets S and D are used. 

Note, however, that this conclusion is based on averaged
quantities. The evolution of local magnetic  
structures is rather complex and possibly also affected by the KH
instability. The radial velocity gradient in Run~T04 is
smaller, so it is expected to have a less efficient KH instability and
this should allow a longer stay of the magnetic field in the stable 
layer. This is hard to demonstrate since it is difficult to disentangle 
these effects in the simulations. Nevertheless, it seems 
that the longer the magnetic field stays at the stable region the
greater the final mean magnetic field strength. Hence, according to
\cite{Fan_01}, the number of scale heights the magnetic flux
concentrations may rise, and its final structure depends only on how
strong the field is in comparison to the turbulent convective
motions.  

In the simulations it is observed that when the magnetic field rises
it expands as a consequence of decreasing density. The morphology 
of the magnetic fields in the $xz$ plane (see bottom panels of
Fig.~\ref{fig:byuz}) is reminiscent of the  
mushroom-shape that has been obtained in several 2D
\citep{schussler_79,moreno+emonet_96,Longcope+etal_96,
  Fan+etal_98,Emonet+Moreno_98}
and 3D \citep{Fan_01,fan+etal_03,Fan_08,jouve+brun_09} simulations of 
flux tube emergence.
Such expansion may result in the 
splitting and braking of the tube. We do not impose
any twist on the tubes since the dynamo generates the magnetic field
self-consistently and the possibilities span from events that do not
rise at all to events where buoyant magnetic fields rise up to the
very surface. In the horizontal $yz$ plane we observe that the
magnetic
field lines remain horizontal in the stable layer where they
form. When the magnetic field rises, the field lines bend
in the convection zone, with their rising part in
the middle of upward flows and other parts withheld to the downward 
flows.
\begin{figure*}[t]
\centering
\includegraphics[width=.99\columnwidth]{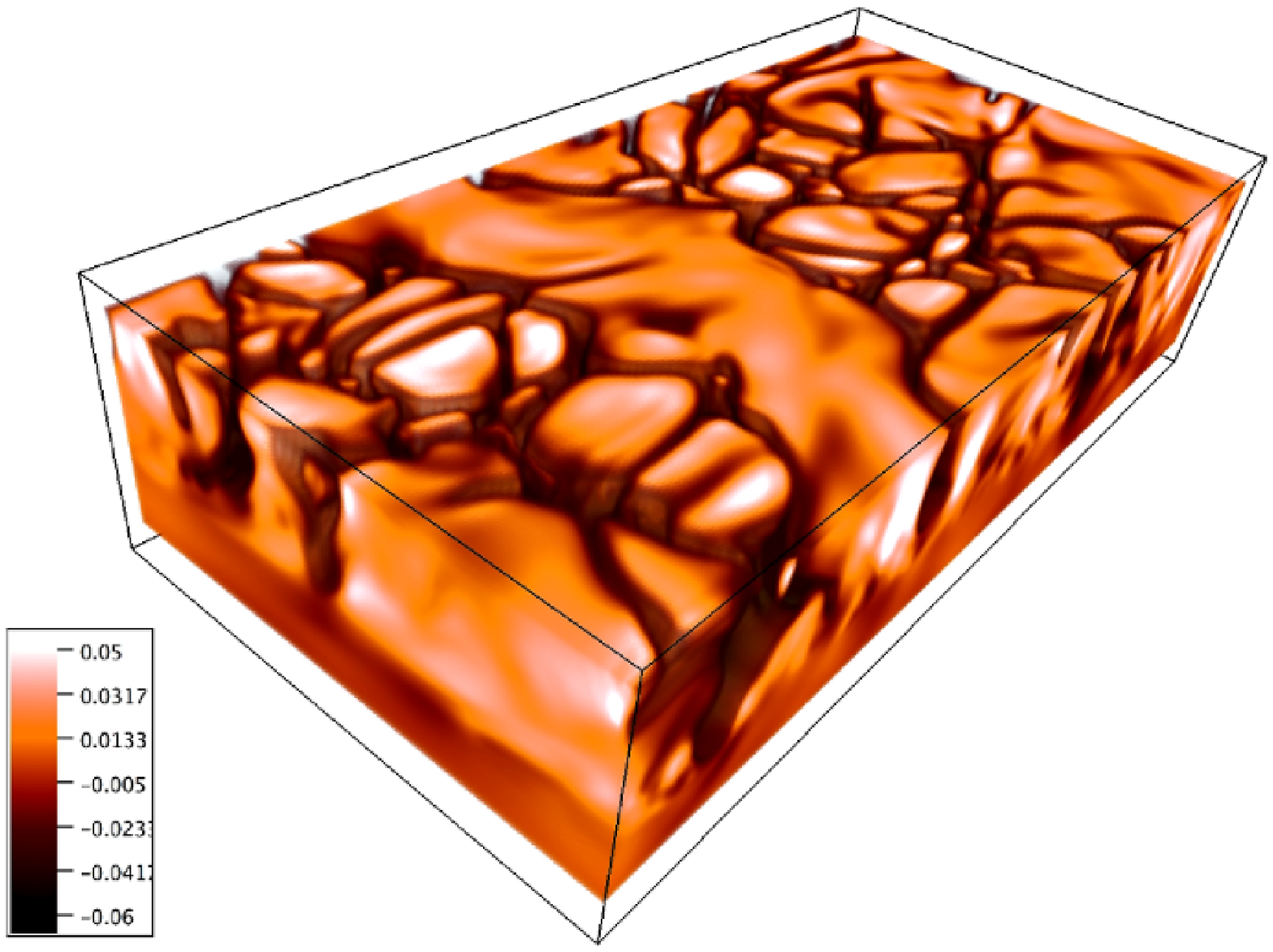}
\includegraphics[width=.99\columnwidth]{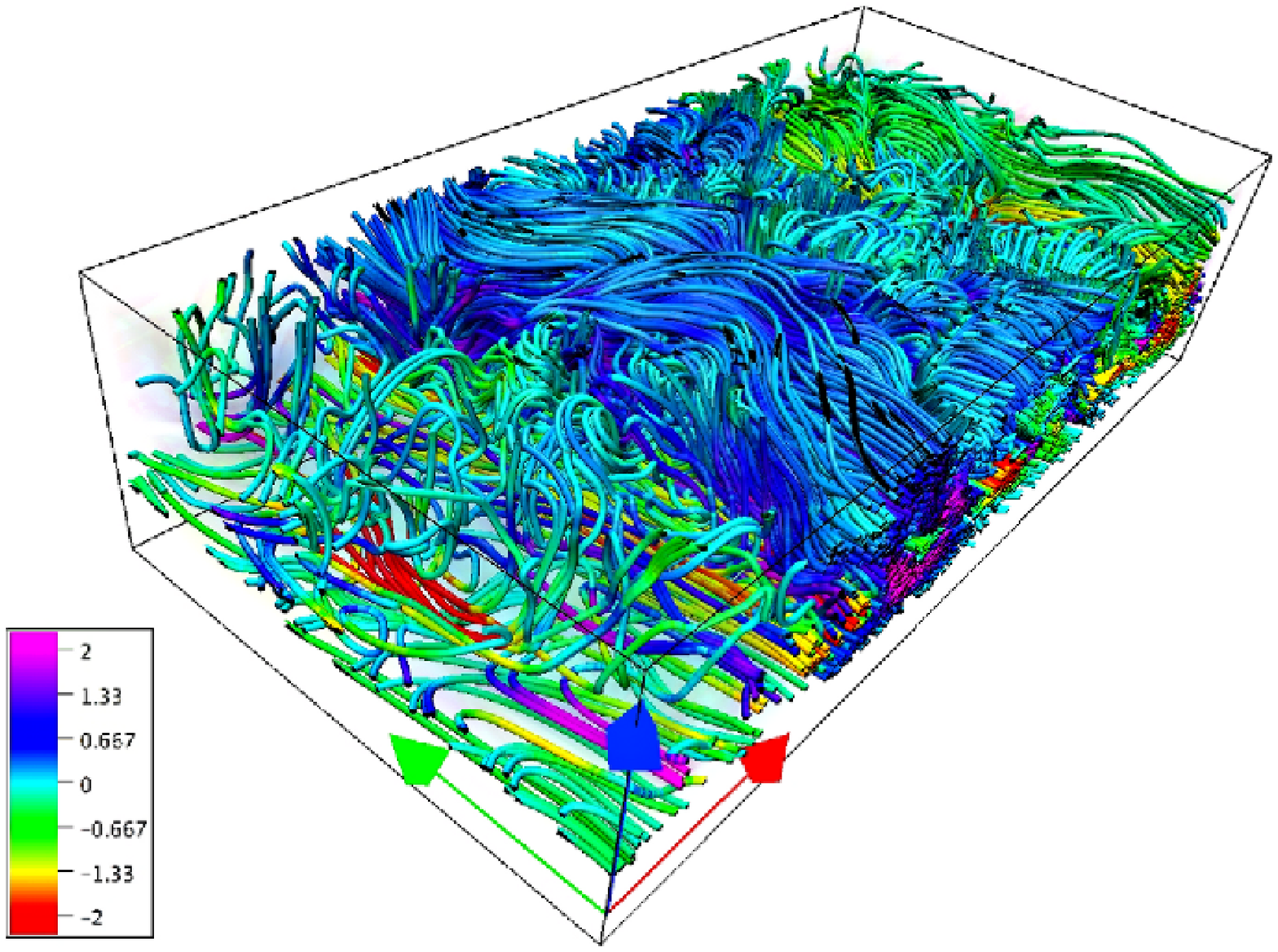}\\
\includegraphics[width=.99\columnwidth]{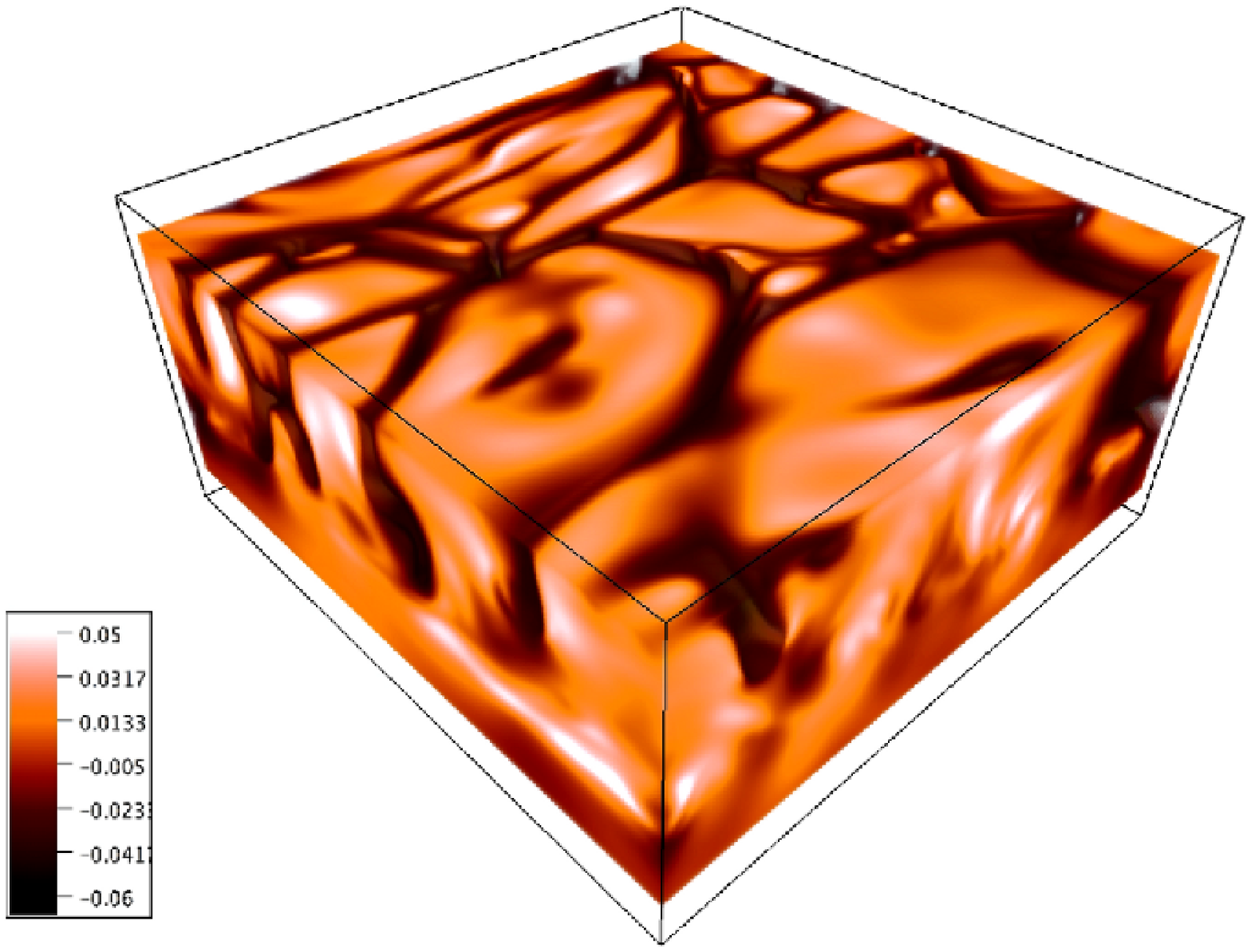}
\includegraphics[width=.99\columnwidth]{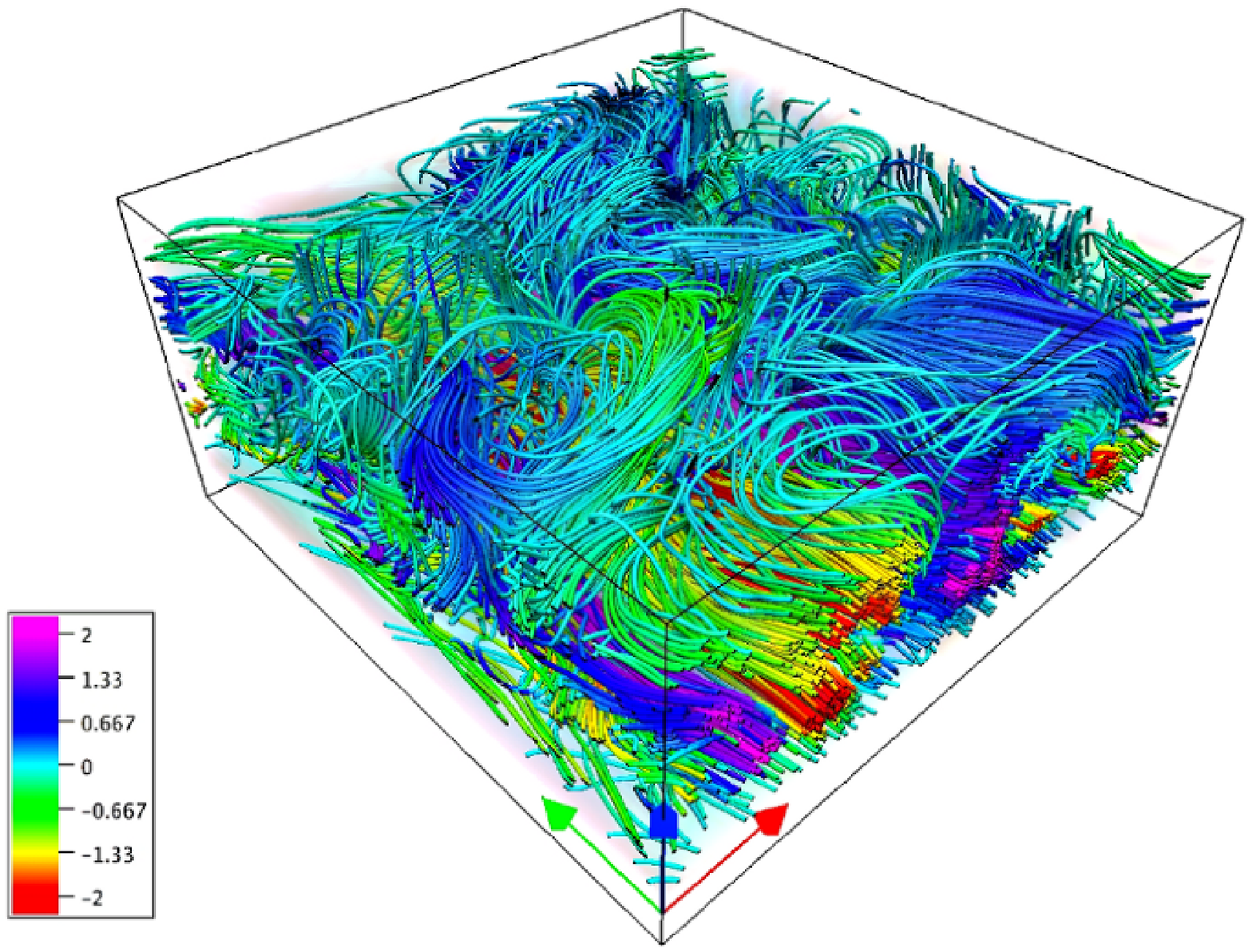}
\caption{Vertical velocity (left panel) and magnetic field lines
  (right panel) for one of the emergence events in
  Runs~AR01 and T04c. The colors of the magnetic field correspond to the values of $B_y/\Beq$.} 
\label{fig:3d}
\end{figure*}

In Tables~\ref{tab:1} and \ref{tab:2} we list the maximum 
values of $B_y$ in the simulations. In most of the cases
we find ${\rm max}(B_y)\approx 5\Beq$, and only in the models with a thicker
(Runs~T04 and AR01) or a deeper shear layer (Run~D04), ${\rm max}(B_y)$ 
can be somewhat greater than $6\Beq$. \cite{fan+etal_03} argue that if 
$B_0 > (H_{\rm P}/a)^{1/2} \Beq$, $B_0$ being the field strength in the
shear layer and $H_{\rm P}$ the local pressure scale height, a buoyant 
flux-tube of radius $a$ may rise without experiencing the 
convective drag force.  Here, although the shape of the magnetic field
is not a tube, we compute the same condition using $a=d_z$. We
obtain values of $B_0$ going from $3.4 \Beq$ in Run~D04 to $2.2\Beq$ in
Run~T04 ($2.3\Beq$ in higher resolution runs).  This 
indicates that the dynamo generated magnetic field is
insufficient to reach the surface without being modified by the
convection. There are, however, a few cases in Runs~D04, T04 
and AR01, where strongest magnetic fields are able to reach the
surface. When this occurs the magnetic field reacts back on the
flow and modifies 
the convective pattern. Broad convection cells elongated in the $y$
direction are formed. In Fig.~\ref{fig:3d} we present one of these
events for Run~AR01. 
In the upper layers, the field
lines show a turbulent pattern except in the center and the 
right edge of the box where the field lines at the surface cross 
the entire domain in the $y$ direction.
In the higher resolution cases, we do not observe events where 
the field lines, in the uppermost layers,
remain horizontal accross the whole toroidal 
direction. Nevertheless, the expanding magnetic field may reach
the surface forming broad convection cells (at least two times
larger than the regular cells). One of 
these cases, corresponding to Run~T04c is shown in the bottom panels of  Fig.~\ref{fig:3d}.
In this event two stripes of magnetic field of opposite polarity are reaching the surface simultaneously, the magnetic field lines of these ropes reconnect forming a large scale loop oriented in the $x$-direction.

\begin{figure}[t]
\centering
\includegraphics[width=.99\columnwidth]{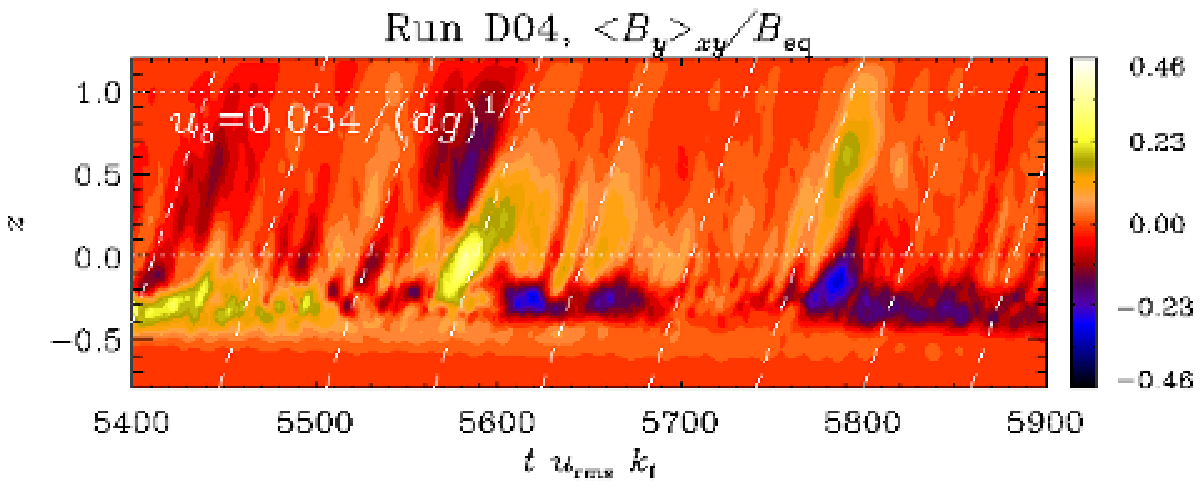}
\includegraphics[width=.99\columnwidth]{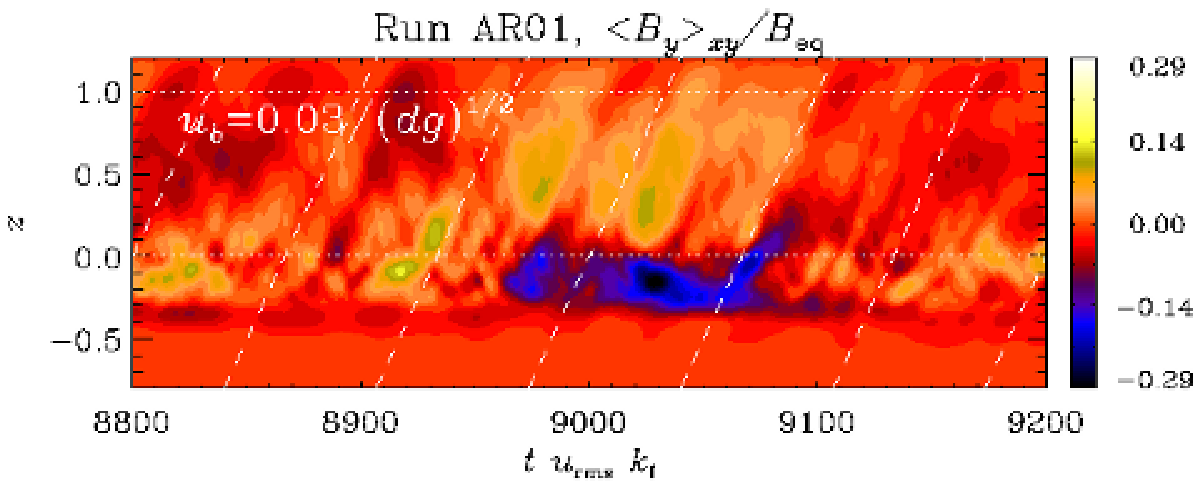}
\caption{Butterfly diagram in the $z$-time plane for the toroidal
  magnetic field, $B_y$. The dotted lines corresponds to the
  top ($z=d$) and base ($z=0$) of the convection zone. The tilted
  dashed lines show approximate trajectories of
  buoyant magnetic fields. A rough estimate of rise speed is computed
  with these lines as indicated. The event at $t=9020 \urms k_f$ in the
  upper panel of Fig. \ref{fig:but} is also shown in Fig. \ref{fig:3d}.} 
\label{fig:but}
\end{figure}

In order to compute the rise speed of the magnetic field in
these cases we have used 
the horizontal averages of $B_y$ and constructed the time-depth
``butterfly'' diagram  shown in Fig. \ref{fig:but}. The
toroidal field is amplified in the shear layer, below $z=0$ (see
dotted line). When it
becomes buoyant it travels through the convection zone.  The tilt
observed in the contours of $B_y$ may give a rough estimate of
the vertical velocity. In the same figure we have drawn white dashed 
lines to guide the eye. For the emergence events in Run~D04 (AR01),
the estimated rise speed is $u_b\approx0.034 (dg)^{1/2}$ ($\approx
0.030(dg)^{1/2}$). 
These values are small in comparison to the $\urms$ of the models
(see Tables~\ref{tab:1} and \ref{tab:2}) which does not agree with the
rise of a flux tube in a stratified atmosphere. For instance,
\cite{fan+etal_03} obtain a final rise velocity $\approx5$ times
larger than the rms-velocity. We should point out that in
the simulations here, the strong shear increases $\urms$ by a factor
of two or even more. If we compare the rise velocity of the magnetic
field with the rms-velocity of a non-shearing case (Run~S00), we find
that $u_b$ is slightly larger. Magnetic pumping effects, which are
evaluated below, may also play a role in braking the magnetic flux and
in defining its final velocity.

The results mentioned above do not fully agree with those found by 
\cite{vasil+etal_08}, where a toroidal magnetic layer is generated from
a purely vertical field through an imposed radial shear layer.
Although their stratification is smaller than in our case, 
the buoyancy instability is hard to excite in their simulations
and the buoyancy events are slowed down rather quickly.  
The lack of a dynamo mechanism, able to 
sustain the magnetic field, is maybe the reason of its rapid 
diffusion.

\subsection{Back reaction}

Temporal analysis of helioseismic data has revealed that the solar
angular velocity varies by around $5$ per cent with respect to its mean
value. This fluctuation pattern, like the sunspots, follows an 11-year
cycle, which suggests that it corresponds to the back reaction
of the magnetic field on the plasma motion
\citep[e.g.][]{basu+antia_03,howe+etal_09}. It is also known that 
the amplitude of the meridional circulation varies with the amplitude
of the magnetic field, being smaller when the cycle reaches its
maximum \citep[see e.g][]{basu+antia_03}. Apart from large scale
effects, the magnetic field may also affect the local properties of
the plasma. This is possible the case in the downflows found by
\cite{hindman+etal_09} nearby the bipolar active regions.

In the simulations presented here, due to the different nature of the
dynamo and to the numerical setup, there are neither periodic
oscillations of the magnetic field nor large-scale circulation. We are
able, however, to distinguish the differences between the plasma
properties in the kinematic and the saturated stages. We also
follow the deviations of the averaged flow occurring during
peaks of the magnetic field amplitude. 

\begin{figure}[t]
\centering
\includegraphics[width=.99\columnwidth]{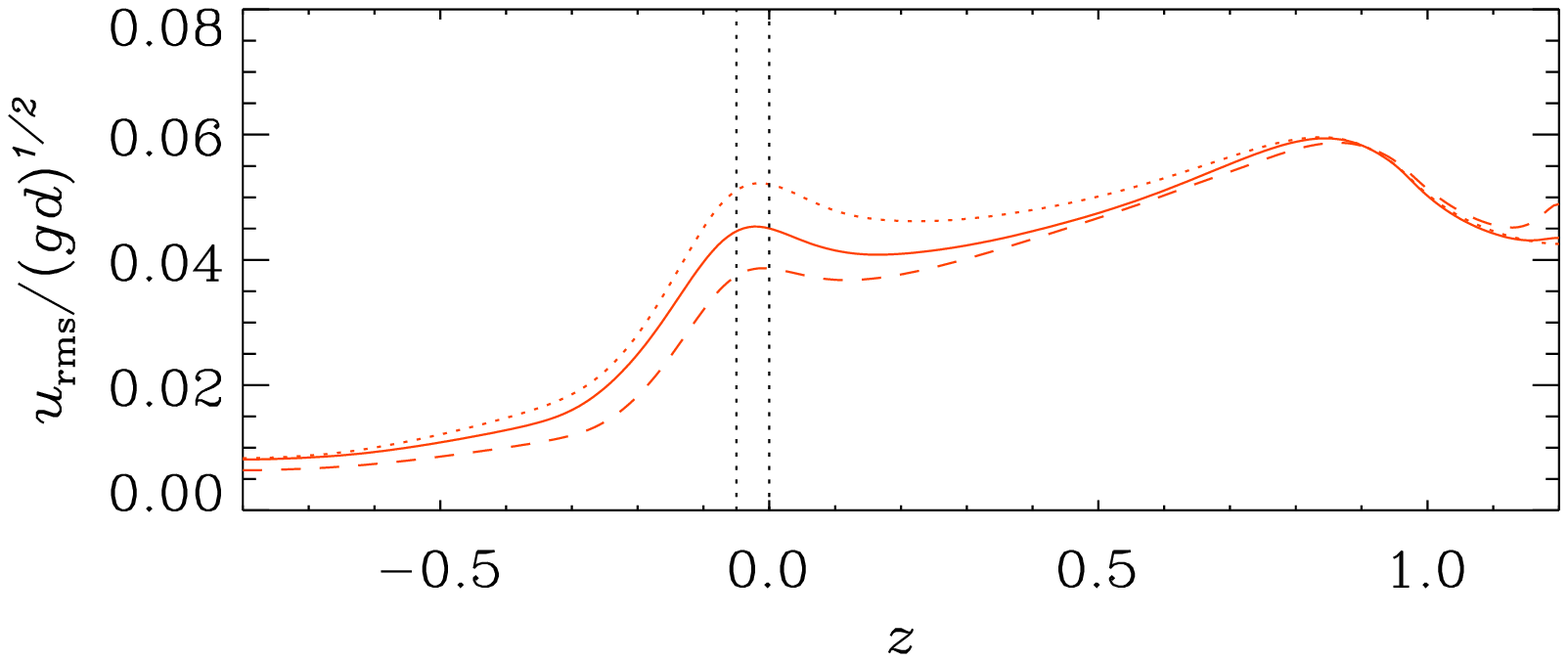}\\
\includegraphics[width=.99\columnwidth]{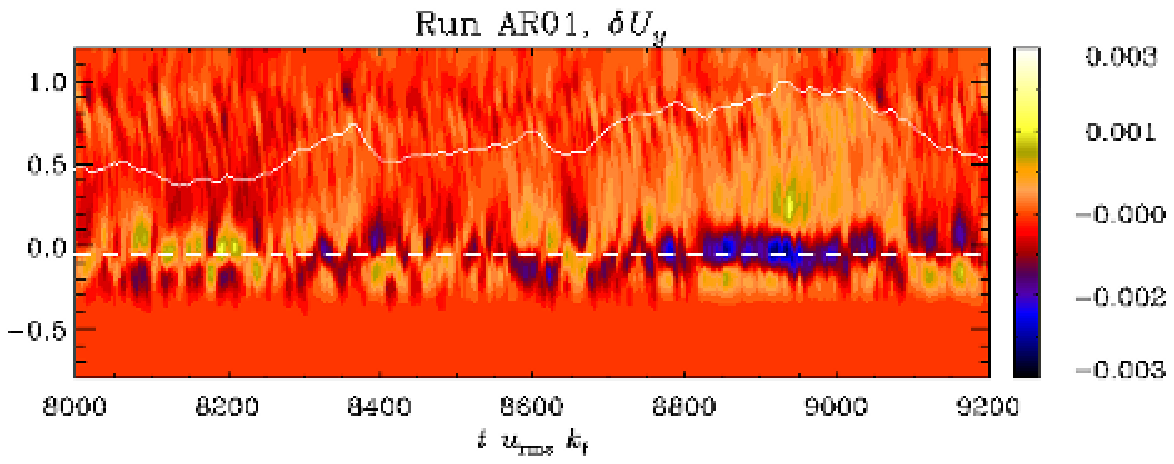}\\
\includegraphics[width=.99\columnwidth]{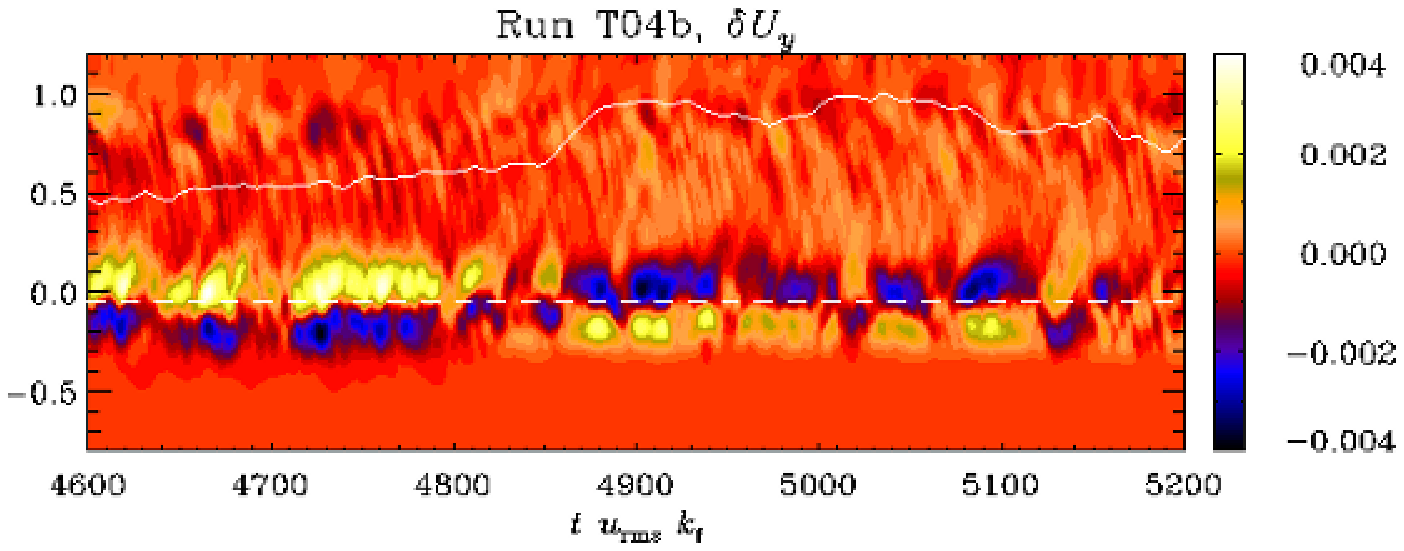}\\
\includegraphics[width=.99\columnwidth]{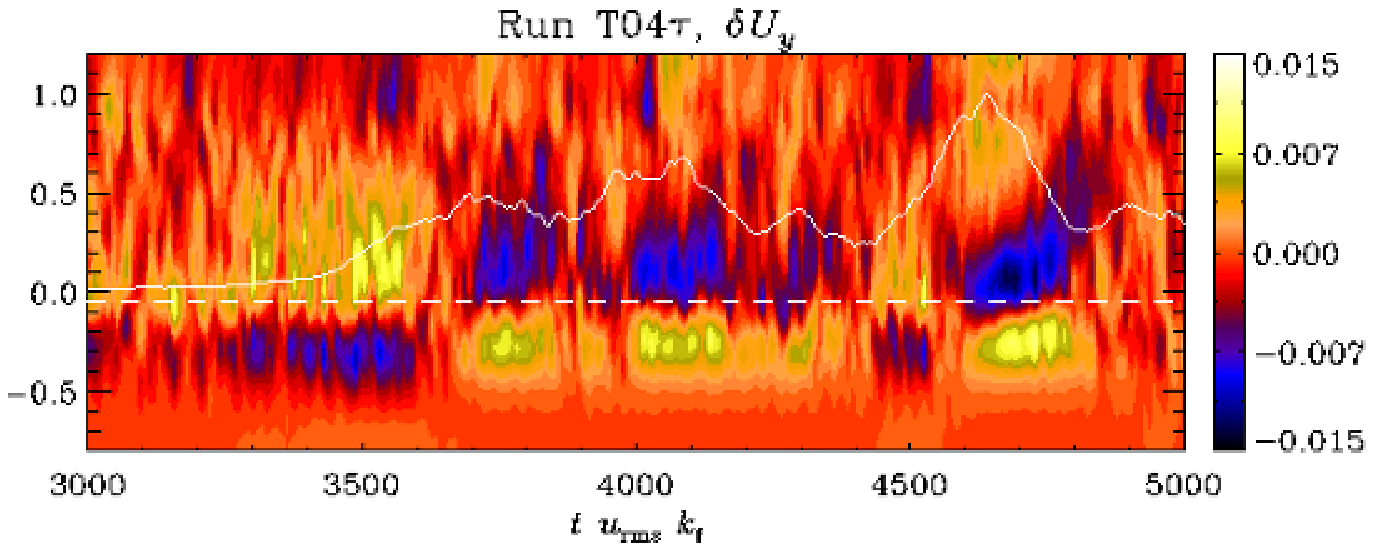}\\
\caption{Top: time averaged vertical profile of $\urms$ in the
  kinematic (dotted line) and dynamical (continuous lined) stages for
  the Runs~T04b. The dashed
  lines show the deviation from the average 
  at a peak in the magnetic field amplitude. The three lower panels
  are butterfly diagrams of the angular velocity deviation, $\delta
  U_y=U_y - \mean{U}_y$, for the Runs~HR01,T04b and T04$\tau$. The
  continuous white line in these panels indicate the time evolution of
  $\brms$ normalized to its maximum value in the time interval.} 
\label{fig:br}
\end{figure}

We find that also the turbulent rms-velocity, as well
as the imposed large-scale shear flow, are affected by the magnetic
field. In the top panel of Fig.~\ref{fig:br} we show the vertical
profile of the horizontal and time averaged $\urms$ at the kinematic
(dotted line) and saturated (continuous line) regimes of the dynamo. A 
decrease in the amplitude is observed, especially in
the region where the magnetic field is more concentrated. The volume
averaged $\urms$ also changes accordingly by $\sim 5$ per cent, as
indicated in Table~\ref{tab:2}. The dashed line in the top
panel of Fig.~\ref{fig:br} shows that it further decreases in the
shear region. 

In the middle and bottom panels of  Fig. \ref{fig:br} butterfly
diagrams of  the toroidal velocity fluctuations,
$\delta U_y=U_y-\mean{U}_y$, 
in the $z$-time plane are shown for the Runs~AR01 and T04b,
respectively. The dashed white line indicates the center of the
shear layer. Positive values of this quantity below
(above) the dashed line indicate faster (slower) toroidal
velocities with respect to its mean value. Accordingly, negative values 
below (above) the dashed line indicate deficiency (excess) of the local $y$
flow. The continuous white line 
corresponds to the time evolution of $\brms$ (for clarity normalized
to its maximum value). We show this in order to
demonstrate that the changes in $U_y$ depend on the amplitude of the 
magnetic field. The overall result of the contour plots is that during
the maximum of the magnetic field the shear is slightly 
stronger. In these diagrams, as well as in all of
the simulations listed in Tables~\ref{tab:1} and \ref{tab:2} (except
Run~T04$\tau$), the deviation around the averaged value of $U_Y$ is
found to be $\sim 3$ per cent.

The deviation of the mean shear profile depends on the forcing
time scale $\tau_{\rm f}$ in
Eq.~\ref{equ:UU}. In Run~T04$\tau$, we consider the same
settings that in Run~T04 but varying $\tau_{\rm f}$ from $2(dg)^{1/2}$ 
to $10(dg)^{1/2}$
($\approx1$ to $\approx5$ turnover times, respectively). The results
show fluctuations of $\approx10$ per cent of the shear velocity, and spread
out now to a wider fraction of the domain (see bottom panel of
Fig.~\ref{fig:br}). Since magnetic buoyancy is present also at the time of the maximum, the toroidal velocity in the middle of the convection zone is also affected. This change is not so strong  as the one in the shear layer but is also a general characteristic of the simulations. 

The perturbations shown in Fig.~\ref{fig:br} are reminiscent of the equatorial branches of the solar torsional oscillations which have a leading excess in  the zonal flow, followed by a slightly  less pronounced deficit of it.
Similar results have also been found in global numerical simulations of oscillating dynamos by \cite{brown+etal_11}. They have found that the branches of zonal flows follow the evolution of a poleward dynamo wave.

\subsection{Turbulent transport coefficients}

In order to study the mechanism that generates a
mean-field dynamo in our simulations,
we compute the turbulent transport
coefficients from a representative simulation (Run~T04). In order to
do this, we use the
test-field method, introduced in the geodynamo context by
\cite{Schrinner+etal_05,Schrinner+etal_07} and currently available in
the {\sc Pencil Code} \citep[e.g.][]{B05,BRRK08}.  

In mean-field theory, the electromotive force
$\memf=\mean{\bm{u}\times\bm{B}}$ governs the evolution of the
large-scale magnetic field \citep[]{KR80}. Under the assumption that
the mean field varies smoothly in space and time, and that there is no
small scale dynamo action, the electromotive force may be written in
terms of the large-scale magnetic field:
\begin{equation}
\mean{\cal{E}}=\alpha_{ij}\mean{B}_j - \eta_{ijk}\mean{J}_k.
\end{equation}
Considering large-scale fields that depend only on $z$ we need four
independent test fields in order to compute the 4+4 components of
$\alpha_{ij}$ and $\eta_{ijk}$ \citep[see a detailed description of
the method in][]{BRRK08}.
The novel feature of the test field method is that the test fields do
not act back on the flow and that the turbulent diffusivity can also
be computed, thus avoiding many of the problems that plaque other
methods \citep[cf.][]{KKB10}.

\begin{figure}[t]
\centering
\includegraphics[width=.99\columnwidth]{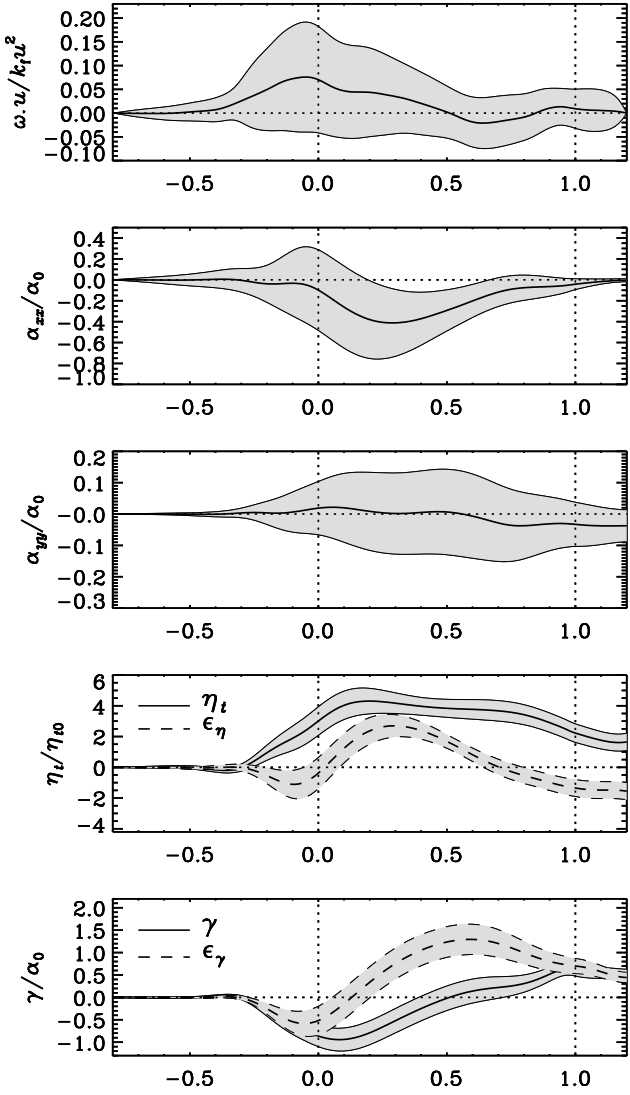} 
\caption{Normalized profiles of the kinetic helicity, and the
  turbulent transport coefficients $\alpha_{xx}$, $\alpha_{yy}$, 
  $\gamma$ and $\etat$ 
  (from top to  bottom) computed with the test-field method.}
\label{fig:tfcoeff}
\end{figure}

We discuss our results in terms of the following 
quantities: 
\begin{eqnarray}
\label{eq:tca}
&\alpha_{xx}& =\alpha_{11}, \quad  \alpha_{yy} = \alpha_{22} ,\\
\label{eq:tce}
&\eta_t &= \onehalf(\eta_{11} + \eta_{22}) , \quad  \epsilon_{\eta} =
\onehalf(\eta_{11} - \eta_{22}),\\
\label{eq:tcg}
&\gamma&=\onehalf(\alpha_{21} - \alpha_{12}), \quad
\epsilon_{\gamma}=\onehalf(\alpha_{21} + \alpha_{12}), 
\end{eqnarray}
which represent the inductive ($\alpha$), diffusive ($\etat$), and
pumping ($\gamma$)
effects of turbulence, respectively. In Fig.~\ref{fig:tfcoeff} we
show the vertical profiles of the kinetic helicity 
($\mean{\bm{u}\cdot\bm{\omega}}$), where $\bm=\bm\nabla\times\bm{u}$ 
is the vorticity, and the turbulent transport
coefficients of Eqs.~(\ref{eq:tca})--(\ref{eq:tcg}), normalized with the
first order smoothing approximation (FOSA) quantities:
\begin{equation}
\alpha_0 = \onethird\urms, \quad \eta_{t0} = \onethird\urms k_{\rm
  f}^{-1}.
\end{equation}
For computing the coefficients in Fig.~\ref{fig:tfcoeff} we use Run~T04 in the purely hydrodynamic state. The magnetic diffusivity considered for the test fields is one order of magnitude larger that that used for the magnetic field in the original run.  However, the magnetic Reynolds number ($\Rm\approx4$) is still sufficiently large to yield reasonable results \citep{Kapyla+etal_09b}.

We find that certain amount of helicity is
generated in the system, but that it is still statistically consistent
with zero. The coefficient $\alpha_{xx}$, despite large
fluctuations, has a negative sign. This component of $\alpha$
contributes to amplifying the $y$ component of the magnetic field, but
its contribution is likely negligible when compared with that 
of the shear in the present models.

The component $\alpha_{yy}$ shows a zero mean value, 
but its variance has a large amplitude, especially in the 
middle of the convection zone. According to \cite{vish+bran_97}, an
incoherent $\alpha$ effect, zero in average but with finite variance,
may generate enough inductive effects to sustain the dynamo.
This suggest that the incoherent $\alpha$ effect may be the
mechanism sustaining the dynamo. 

The shear-current effect, arising from the inhomogeneity of the
turbulence and the mean shear flow, may also result in the generation
of a mean field \citep{rog+kleo_03,rog+kleo_04}. The existence of this
effect has been studied with models of forced turbulence and
latitudinal shear by \citep{BRRK08, Mitra+etal_09}, but no conclusive
evidence has been found. Similar results have also been found from
convection with horizontal shear \citep{Kapyla+etal_09b}. 
In the case presented here,  with vertical shear, the relevant
coefficient for this effect would be $\eta_{13}$, which is not
captured by the current version of the test-field method with only
$z$-dependent test fields. Thus, we
leave the study of the shear-current effect in turbulent convection
with vertical shear for a forthcoming work.  

The turbulent diffusivity, $\eta_t$, is roughly four times larger than the
FOSA estimate. When compared with the molecular diffusivity,
it is also four times larger (i.e.\ of the same order as
$\Rm \sim \etat/\eta$). This result agrees with
previously computed turbulent diffusivity for convection 
\citep{Kapyla+etal_09b}. The
vertical profile of $\etat$ differs 
somewhat from previous results, indicating the action of the
radial shear on the turbulent diffusion.  Another interesting point is
that $\epsilon_{\eta}$ is not close to zero, as in
\cite{Kapyla+etal_09b}. This is likely explained by the stronger shear 
used here. This is also an indication of the tensorial
character of $\etat$ and suggests  that the distinct 
components of the magnetic field may diffuse differently.

Finally, the vertical profile of the turbulent pumping ($\gamma$) in
Fig. \ref{fig:tfcoeff} shows that there is downwards transport of
the mean magnetic field at the bottom of the convection zone and
upward transport close to the top.  The  pumping velocity is 
comparable with $\urms$ but it is not enough to retain the
magnetic field within the stable layer, which indicates that buoyancy is
a more efficient mechanism than magnetic pumping in transporting
magnetic fields. Note, however, that the amplitude of this effect
depends on the gradient of the level of turbulence, $\gamma = -\nabla 
\eta_{\rm T}$, where $\eta_{\rm T} = \eta + \etat$ \citep{Kit+Rud_92}, which is
not very large in the current simulations. 


\section{Conclusions}
\label{sec:conclusions}

We perform numerical simulations of turbulent convection with a thin
vertical shear layer with the aim of mimicing the conditions at the
solar convection zone and the tachocline. This layer is located
below the interface between a convectively stable and
unstable layers. We consider the velocity gradient, the
depth and the thickness of this layer as free parameters of the
model and concentrated on two phenomena relevant for the solar
cycle: dynamo process and magnetic buoyancy.

We find that it is possible to excite a large scale
dynamo by the combined action of turbulent convection and a localized 
radial shear. The efficiency of the dynamo (i.e., the growth
rate), and the maximum amplitude of the magnetic field depend on
the shear amplitude, the thickness of the shear layer, the aspect
ratio and the magnetic Reynolds number of the system (see
Fig.~\ref{fig:b-t} and Tables~\ref{tab:1} and \ref{tab:2}).

The magnetic field is organized in elongated structures in the
direction of the shear in the shear layer below the base of the
convection zone. The large-scale
magnetic field can have either polarity and may be up to the $40$ per 
cent of the total field (solid lines in Fig.~\ref{fig:fvsm}). It coexists with
turbulent magnetic field which is distributed all across the
convection zone. In the stable 
layer the magnetic field exists in the regions of strong downflows. 
For fixed depth and thickness (Set~S) a smaller growth
rate and larger $\brms$ is obtained for a larger shear parameter. A
critical value of $\rm{Sh}\approx7$ is required to excite the dynamo.

If the tachocline is located deeper (simulations in set D), the
magnetic field develops mainly in the stable layer which allows a
longer field storage. In 
these cases the dynamo grows faster and with a larger fraction of mean 
field magnetic field. This configuration, however, is 
not optimal since with the numerical resolution used here the
thermodynamical properties of the fluid are affected by viscous
heating in the shear region. If the depth of the shear layer remains
constant but the thickness increases (Set~T), the
magnetic field grows slowly but it contains a larger fraction of mean
magnetic field. This may be a consequence of the smoother
shear profile which renders the magnetic field buoyantly unstable only
in the superadiabatic part of the domain (see right panel of
Fig.~\ref{fig:bi}). A lower critical shear number is found to be able
of excite the dynamo instability in these cases ($\rm{Sh} \approx
3.5$).

Larger magnetic Reynolds numbers
are achieved in two ways, first by doubling the initial resolution and
second by changing the input flux (i.e., different $\Pra$ number). On
one side, for a fixed $\Pra$ and larger $\Rm$, the dynamo exhibits a
much faster growth. On the other hand, for a fixed resolution and
different input flux, the growth rate changes proportionally to
$\Pra$. In these higher resolution Runs the final value of $\brms$ is
slightly below that in the corresponding cases with lower
resolution. This may correspond to the dependence of the saturation
value of the magnetic field with the magnetic Reynolds number or may
also be the result of insufficient statistics.

Since the system is not rotating, there is no kinetic helicity and
hence no $\alpha$-effect. The test-field analysis
suggests that the
probable mechanism triggering the amplification
of the magnetic field could be incoherent $\alpha$-shear
dynamo. However, other possibilities like the shear-current dynamo can 
not be ruled out for the time being.

Magnetic buoyancy is observed in all of the simulations. Based on
horizontal averages we analyze the instability condition for 2D
interchange modes (Eq.~\ref{eq:binst} and Fig.~\ref{fig:bi}). We 
find that in models 
with a deeper tachocline the buoyancy instability develops even in the
stable layer whereas in models with thicker shear layers the magnetic
field is unstable in the convection zone only. This, however, does
not mean that there are not emergence events. As far as the toroidal
magnetic field is strong enough it rises throw the convection zone
forming mushrooms shape structures.

When the buoyant magnetic field is
weak it is strongly modified by local convective flows (see
Fig.~\ref{fig:byuz}). On the other hand, magnetic fields of strong
amplitude formed at the shear 
layer are able to rise up to the surface and modify the convective
pattern (Fig.~\ref{fig:3d}). They may form either large convection cells
or convective rolls may occupy all of the $y$-extent.
These clearest events are 
observed when the magnetic field in the shear layer exceeds the
equipartition field by a factor $\geq 6$. Such strong magnetic field
has been observed only in simulations with $128^3$ grid points
resolution. The rise velocity of the buoyant field in the bulk of
the convection zone, estimated from $z$-time ``butterfly'' diagrams,
has been found that $u_b \approx 0.6 \urms$. From the test-field method
results, the maximal downwards pumping velocity of magnetic field is
found to be $\gamma \approx 0.4 \urms$. This indicates that buoyancy may
indeed exist from the base of the convection zone, however, the
magnetic field expands very quickly during the rise and 
magnetic structures observed close to the upper boundary are not
so well organized. 

Besides the emergence of the magnetic field that affects the flow
pattern locally, other changes are observed due to the back-reaction
of the magnetic force on the plasma. In the saturated phase, the 
rms-velocity varies between its kinematic value and a lower amplitude
when the magnetic field is strong. In addition, during the peaks and  
valleys of the field amplitude, the imposed shear profile
presents systematic variations with respect to its averaged profile. 
For most of the models this
change is around $3$ per cent of the shear velocity and is
reminiscent of the fluctuations observed in the solar rotation
profile or ``torsional oscillations''. The variation
signal is mainly observed in the places where ${\bm B}$ is strong,
with much weaker changes within the convection zone. The results
encourage the study of the torsional oscillations in detail 
through direct numerical simulations.

\begin{acknowledgements}
  The authors acknowledge the hospitality of NORDITA.
  This work is supported by the European Research Council under the
  AstroDyn research project 227952. The computations were performed
  under the HPC-EUROPA2 project (project number: 228398) with the
  support of the European Commission -- Capacities Area -- Research
  Infrastructures. PJK acknowledges the financial support from the
  Academy of Finland grant Nos.\ 121431, 136189, and 140970.
\end{acknowledgements}

\bibliographystyle{aa}
\bibliography{bib}

\end{document}